\documentclass[11pt,english,longbibliography]{article}
\usepackage{jheppub}
\usepackage{./jheppub}
\usepackage[utf8]{inputenc}
\usepackage[english]{babel}
\usepackage[T1]{fontenc}
\usepackage[titletoc]{appendix}
\usepackage{titletoc}
\usepackage{graphicx}
\usepackage{subcaption} %
\usepackage{xcolor}
\usepackage{amsmath,amssymb,amsthm}
\usepackage{bm}
\usepackage{float}
\usepackage[format=plain,labelfont={bf,it},textfont=it]{caption}
\usepackage{multirow}
\usepackage{slashed} %
\usepackage{tabularx}
\usepackage{bbold}
\usepackage[shortlabels]{enumitem}
\usepackage{tikz}
\usetikzlibrary{positioning,arrows,calc}
\usetikzlibrary{arrows.meta}
\usetikzlibrary{decorations.pathmorphing}
\usetikzlibrary{decorations.markings}
\usetikzlibrary{shapes.geometric}
\usetikzlibrary{patterns}
\tikzset{
  snake it/.style={
    decorate, 
    decoration=snake,
    segment length=3
  }
}
\usepackage{xparse}
\usepackage{pdfpages} %
\usepackage{titlesec} %
\usepackage{fancyhdr} %
\usepackage{emptypage} %
\usepackage{dsfont} %
\usepackage{bm}
\usepackage{graphicx}
\usepackage{caption}
\usepackage{multirow}
\usepackage{subcaption}
\usepackage{ragged2e}
\usepackage{cancel}
\usepackage{comment}
\usepackage{xcolor}
\DeclareCaptionJustification{justified}{\justifying}
\usepackage{hyperref}
\usepackage{amssymb}
\usepackage{amsmath}

\usepackage{tikz}
\usetikzlibrary{calc,arrows.meta}

\newcommand{\veps}{\varepsilon}

\definecolor{DarkBlueGrey}{RGB}{76,94,107}
\definecolor{MediumBlueGrey}{RGB}{110,135,153}
\definecolor{LightBlueGrey}{RGB}{134,163,184}
\definecolor{WCOrange}{RGB}{242,146,29}
\definecolor{SCRed}{RGB}{179,48,48}
\definecolor{SCRed2}{RGB}{230,100,100}
\definecolor{VertexColor}{RGB}{242,146,29}
\definecolor{GluonColor}{RGB}{255,172,172}
\definecolor{SEColor}{RGB}{134,163,184}

\definecolor{BGBox}{RGB}{255,254,230}
\definecolor{PlaneColor}{RGB}{230,230,230}
\definecolor{BlobColor}{RGB}{190,180,230}

\def\veps{\varepsilon}

\usepackage{bbm}

\def\veps{\varepsilon}

\newif\ifstartcompletesineup
\newif\ifendcompletesineup
\pgfkeys{
    /pgf/decoration/.cd,
    start up/.is if=startcompletesineup,
    start up=true,
    start up/.default=true,
    start down/.style={/pgf/decoration/start up=false},
    end up/.is if=endcompletesineup,
    end up=true,
    end up/.default=true,
    end down/.style={/pgf/decoration/end up=false}
}
\pgfdeclaredecoration{complete sines}{initial}
{
    \state{initial}[
        width=+0pt,
        next state=upsine,
        persistent precomputation={
            \ifstartcompletesineup
                \pgfkeys{/pgf/decoration automaton/next state=upsine}
                \ifendcompletesineup
                    \pgfmathsetmacro\matchinglength{
                        0.5*\pgfdecoratedinputsegmentlength / (ceil(0.5* \pgfdecoratedinputsegmentlength / \pgfdecorationsegmentlength) )
                    }
                \else
                    \pgfmathsetmacro\matchinglength{
                        0.5 * \pgfdecoratedinputsegmentlength / (ceil(0.5 * \pgfdecoratedinputsegmentlength / \pgfdecorationsegmentlength ) - 0.499)
                    }
                \fi
            \else
                \pgfkeys{/pgf/decoration automaton/next state=downsine}
                \ifendcompletesineup
                    \pgfmathsetmacro\matchinglength{
                        0.5* \pgfdecoratedinputsegmentlength / (ceil(0.5 * \pgfdecoratedinputsegmentlength / \pgfdecorationsegmentlength ) - 0.4999)
                    }
                \else
                    \pgfmathsetmacro\matchinglength{
                        0.5 * \pgfdecoratedinputsegmentlength / (ceil(0.5 * \pgfdecoratedinputsegmentlength / \pgfdecorationsegmentlength ) )
                    }
                \fi
            \fi
            \setlength{\pgfdecorationsegmentlength}{\matchinglength pt}
        }] {}
    \state{downsine}[width=\pgfdecorationsegmentlength,next state=upsine]{
        \pgfpathsine{\pgfpoint{0.5\pgfdecorationsegmentlength}{0.5\pgfdecorationsegmentamplitude}}
        \pgfpathcosine{\pgfpoint{0.5\pgfdecorationsegmentlength}{-0.5\pgfdecorationsegmentamplitude}}
    }
    \state{upsine}[width=\pgfdecorationsegmentlength,next state=downsine]{
        \pgfpathsine{\pgfpoint{0.5\pgfdecorationsegmentlength}{-0.5\pgfdecorationsegmentamplitude}}
        \pgfpathcosine{\pgfpoint{0.5\pgfdecorationsegmentlength}{0.5\pgfdecorationsegmentamplitude}}
}
    \state{final}{}
}

\tikzset{
corner/.style={line width=1pt,dashed,draw=black,dash pattern=on 6pt off 4pt},
scalar/.style={line width=1pt,draw=black},
gluon/.style={line width=1pt,decorate, draw=GluonColor,
    decoration={complete sines,aspect=0,amplitude=1.25mm,segment length=1.5mm,start up,end up}},
gluontwo/.style={line width=1pt,decorate, draw=GluonColor,
    decoration={complete sines,aspect=0,amplitude=.7mm,segment length=1mm,start up,end up}},
ghost/.style={line width=1pt,loosely dotted,draw=black},
wilson/.style={line width=2pt,draw=black},
 }
\NewDocumentCommand\semiloop{O{black}mmmO{}O{above}}
{%
\draw[#1] let \p1 = ($(#3)-(#2)$) in (#3) arc (#4:({#4+180}):({0.5*veclen(\x1,\y1)})node[midway, #6] {#5};)
}

\usepackage{indentfirst}
\usepackage{yfonts}
\usepackage{enumitem}
\usepackage{pgfplots}
\pgfplotsset{compat=1.18}
\usepgfplotslibrary{fillbetween}
\usetikzlibrary{patterns, positioning, fadings, arrows.meta, calc,decorations.text}
\title{An Effective Theory for Hot Lines}

\author[a]{Julien Barrat,}
\author[b]{Alessio Miscioscia,}
\author[c]{and Fedor K. Popov}

\affiliation[a]{Department of Physics, Princeton University, Princeton, NJ 08544, USA}
\affiliation[b]{C. N. Yang Institute for Theoretical Physics, Stony Brook University, Stony Brook, NY 11794,USA}
\affiliation[c]{Simons Center for Geometry and Physics, Stony Brook University, Stony Brook, NY}
\abstract{

We develop a high-temperature effective theory for spatial conformal line defects in theories with thermal screening.
The defect partition function admits a local geometric expansion, whose leading terms describe the effective tension and bending response of the line.
Using a convergent bulk OPE for heavy-state defect matrix elements, we derive asymptotic relations for averaged products of defect one-point coefficients and bulk OPE coefficients in the regime $1 \ll \Delta_\chi \ll \Delta^{1/d}$.
We compute the leading EFT coefficients for a twisted free scalar and for the magnetic line at the $\mathrm{O}(N)$ Wilson--Fisher fixed point, large-$N$ Gross-Neveu model and $\mathcal N = 2$ Chern-Simons-matter theory.
For the three-dimensional Ising model, Monte Carlo simulations yield a perimeter coefficient in good agreement with a Pad\'e interpolation of our $\varepsilon$-expansion result.
}

\begin{document} 

 \vspace*{-.6in}
 \begin{flushright}
     YITP-SB-2026-18
 \end{flushright}

\maketitle

\section{Introduction and summary}
\label{sec:IntroductionAndSummary}

Conformal field theories (CFTs) describe the universal behavior of systems at continuous phase transitions and, more generally, fixed points of renormalization-group flows~\cite{WilsonKogut1974,DombGreen1976,Amit1984,Cardy1996,Kadanoff2000}.
Their correlation functions are encoded in the conformal data: scaling dimensions and operator product expansion (OPE) coefficients.
Considerable progress has been made in determining and constraining these data through numerical and analytic bootstrap methods~\cite{El-Showk:2012cjh,Fitzpatrick:2012yx,Komargodski:2012ek,El-Showk:2014dwa,Caron-Huot:2017vep,Simmons-Duffin:2017nub,Poland:2018epd,Anand:2025mfh,Komargodski:2026ain,Advant:2026cqt}.

Beyond local correlation functions, one can study finite-temperature observables, such as the free energy and thermal correlation functions.
Although these are also determined by the conformal data, their dependence on the full spectrum and OPE coefficients makes them difficult to compute directly, motivating complementary approaches~\cite{Katz:2014rla,Iliesiu:2018fao}.
Conversely, thermal observables provide access to asymptotic properties of conformal data, including the growth of the density of states described by Cardy-type formulas~\cite{Cardy:1986ie,Shaghoulian:2015kta,Benjamin:2023qsc} and the behavior of heavy-state three-point coefficients~\cite{Buric:2025uqt,Buric:2026pes,Miscioscia:2026bvl}.

A useful approach is to place the theory on a thermal background $S^1_\beta \times \mathcal{M}_{d-1}$, where $\beta = 1/T$ is the circumference of the Euclidean time circle and $\mathcal{M}_{d-1}$ is the spatial manifold.
Thermal periodicity, encoded in the Kubo--Martin--Schwinger (KMS) condition, constrains correlation functions and provides the basis for thermal bootstrap methods~\cite{Iliesiu:2018zlz,Marchetto:2023xap,Barrat:2025wbi,Barrat:2025nvu,Barrat:2026jfg,Guo:2026xyl}.
In the high-temperature limit, $\beta$ becomes small compared with the characteristic spatial scales, and locality allows the leading terms of the partition function to be expressed in terms of local geometric data~\cite{BhattacharyyaEtAl2008,BanerjeeEtAl2012,JensenEtAl2012,Shaghoulian2016,Benjamin:2023qsc}.

Conformal defects introduce additional observables, such as defect partition functions and correlation functions, whose description requires data beyond those of the bulk CFT.
These include the spectrum of defect operators and bulk-to-defect OPE coefficients~\cite{Billo:2016cpy}.
Bootstrap, perturbative, and effective field theory methods have provided complementary results for conformal defects~\cite{Billo:2013jda,Gliozzi:2015qsa,Lemos:2017vnx,Cuomo:2021rkm,Cuomo:2021kfm,Barrat:2022psm,Diatlyk:2024qpr,Meineri:2025qwz,Bianchi:2026dhj}.
More recently, conformal defects have also been studied at finite temperature~\cite{Barrat:2024aoa,Giombi:2026kdz,Li:2026udh,Artico:2026rye,Barrat:2026ozj}, but the relation between their thermal behavior and zero-temperature conformal data remains less understood.

\bigskip

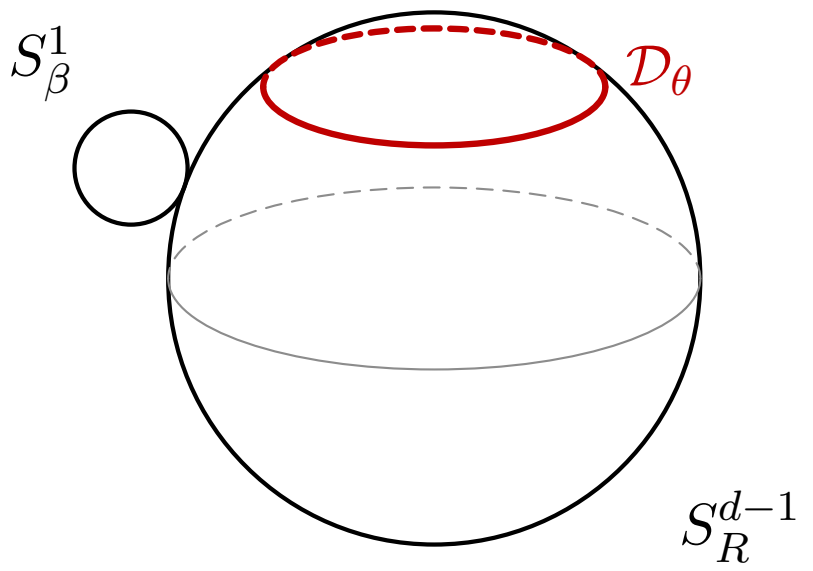
\begin{figure}[t]
    \centering
    \begin{tikzpicture}[scale=1.1, line cap=round, line join=round]
        \def\radius{1.6}
        \def\defectangle{40}
        \def\viewtilt{20}

        \pgfmathsetmacro{\xD}{\radius*sin(\defectangle)}
        \pgfmathsetmacro{\yD}{\radius*cos(\defectangle)*cos(\viewtilt)}
        \pgfmathsetmacro{\yRing}{\xD*sin(\viewtilt)}
        \pgfmathsetmacro{\yEquator}{\radius*sin(\viewtilt)}
        \pgfmathsetmacro{\fiberDistance}{\radius+0.34}

        \coordinate (C) at (0,0);

        \draw[thick] (C) circle (\radius);

        \coordinate (T) at ($(C)+(160:\fiberDistance)$);
        \draw[thick] (T) circle (0.34);
        \node[above left] at ($(T)+(-0.22,0.28)$) {$S_\beta^1$};
        \node[below right] at (1.35,-1.15) {$S_R^{d-1}$};

        \draw[black!45,thin,densely dashed]
            ($(C)+(\radius,0)$)
            arc[start angle=0,end angle=180,
                x radius=\radius,y radius=\yEquator];

        \draw[black!45,thin]
            ($(C)+(-\radius,0)$)
            arc[start angle=180,end angle=360,
                x radius=\radius,y radius=\yEquator];

        \draw[red!75!black,very thick,densely dashed]
            ($(C)+(\xD,\yD)$)
            arc[start angle=0,end angle=180,
                x radius=\xD,y radius=\yRing];

        \draw[red!75!black,very thick]
            ($(C)+(-\xD,\yD)$)
            arc[start angle=180,end angle=360,
                x radius=\xD,y radius=\yRing];

        \node[red!75!black,right]
            at ($(C)+(\xD,\yD+0.10)$)
            {$\mathcal{D}_\theta$};
    \end{tikzpicture}

    \caption{Spatial line defect $\mathcal{D}_\theta$ supported on a latitude at polar angle $\theta$ of the sphere $S_R^{d-1}$, with Euclidean time compactified on $S_\beta^1$.}
    \label{fig:finite-temperature-defect-setup}
\end{figure}

In this paper, we develop a high-temperature effective theory for conformal line defects supported on spatial curves.\footnote{Related effective descriptions have been developed for the fusion of conformal defects and for conformal boundaries~\cite{Diatlyk:2024zkk,Diatlyk:2024qpr,Kravchuk:2024qoh}.
Our setup instead concerns a single spatial defect in a thermal background, with the effective expansion controlled by thermal screening.}
We focus on the spherical geometry illustrated in Figure~\ref{fig:finite-temperature-defect-setup}, where the defect lies at fixed Euclidean time rather than wrapping the thermal circle.
At high temperature, the bulk admits an effective description in $d-1$ dimensions.
Provided the thermal response to the defect is screened, its contribution to the partition function can be organized in a local geometric expansion along $\mathcal{D}$:\footnote{This expansion need not apply to all line defects.
For example, spatial Wilson loops in large-$N$ gauge theories can exhibit an area law at finite temperature as in certain holographic theories~\cite{Maldacena:1998im,Witten:1998zw,Berenstein:1998ij,Alday:2009yn}.
Such contributions cannot be captured by a local effective action along the line.
Notice that this phenomenon is not possible in $\mathcal N = 4$ SYM at finite $N$ since the confinement/deconfinement transition is forbidden on the sphere.}
\begin{equation}
    \log \frac{Z[\beta,\mathcal{D}]}{Z[\beta]}
    = \gamma[\mathcal{D}]
    + \int_{\mathcal{D}} ds \left[
        \frac{m_D}{2\pi\beta}
        + \frac{\beta}{2\pi}
        \left(t_2 K_i K^i+\frac{t_R}{R^2}\right)
        + \ldots
    \right]\,.
    \label{eq:LogPartitionFunctionintro}
\end{equation}
Here, $\gamma[\mathcal{D}]$ denotes the temperature-independent contribution, while $m_D$ and $t_2$ control the effective tension and bending response of the line, respectively.
$t_R$ describes the combined ambient-curvature coefficient.
The omitted terms involve higher-order geometric invariants, including contributions from the curvature of the spatial manifold.

Under the assumptions detailed in Section \ref{sec:asymptotic}, we further show that the thermal EFT determines the asymptotic behavior of averaged products of defect one-point coefficients and bulk OPE coefficients.
The derivation relies on a bulk OPE representation of heavy-state defect matrix elements, building on the known convergence properties of Euclidean OPE expansions~\cite{Pappadopulo:2012jk,Lauria:2017wav}.
We obtain
\begin{equation}\label{eq:AsymptoticRed}
    \overline{a_\chi f_{\mathcal O_\Delta \mathcal O_\Delta \chi}}
    \sim \frac{e^{\widehat\gamma}m_D^{\Delta_\chi}}{\Gamma(\Delta_\chi+1)}
    \left(
        \frac{\Delta}{-(d-1)f}
    \right)^{\Delta_\chi/d}\,,
\end{equation}
in the regime $1 \ll \Delta_\chi \ll \Delta^{1/d}$.
Here $a_\chi$ is the defect one-point coefficient of an operator $\chi$ of dimension $\Delta_\chi$, and $f_{\mathcal O_\Delta \mathcal O_\Delta \chi}$ is its bulk OPE coefficient with two heavy operators of dimension $\Delta$.
We have also defined $\widehat\gamma=\gamma-\log g_D$, where $g_D$ is the zero-temperature defect expectation value, while $f$ denotes the coefficient of the leading high-temperature bulk free energy in the absence of the defect.

This result also admits a thermal interpretation under the eigenstate thermalization hypothesis (ETH).
If the heavy states created by $\mathcal O_\Delta$ are typical, their heavy-heavy-light OPE coefficients can be related to thermal one-point functions.
Under these assumptions, we obtain up to kinematic factor 
\begin{equation*}
    \overline{a_\chi b_\chi}
    \simeq
    \frac{e^{\widehat{\gamma}}m_D^{\Delta_\chi}}
    {\Gamma(\Delta_\chi+1)}\,.
\end{equation*}
where $b_\chi$ denotes the thermal one-point coefficient of $\chi$.

We illustrate the effective field theory (EFT) in a twisted free scalar theory and compute its leading coefficients for the magnetic line at the $\mathrm{O}(N)$ Wilson--Fisher fixed point in the $\varepsilon$-expansion, the large $N$ Gross-Neveu model and the $\mathcal N = 2$ Chern-Simons matter theory. The latter is particular interesting since, differently from the others, the leading EFT coefficient $m_D$ is positive. 
We also study the case of the three-dimensional Ising model and determine the value of the leading EFT coefficient $m_D$ using Monte Carlo simulations and find good agreement with a Pad\'e interpolation of our perturbative results. A recap of the results for $m_D$ in the various model is provided in Table \ref{tab:MDS}
\begin{table}[t]
    \centering
    \caption{Prediction for the leading EFT coefficient $m_D$ in various models. $h$ indicated the conformal coupling of the defect (when indicated the model enjoy a continuum of conformal defects varying the coupling $h$). }
    \begin{tabular}{c c c}
        \hline
        Model & $m_D$ & method \\
        \hline
        Antiperiodic Free Scalar &$- h^2 \ 0.346\ldots $ &Exact  \\ 
        $\mathrm O(N)$ model in $d= 4-\varepsilon$ & eq. \eqref{eq:epsilonexpresult} & Perturbative in $\varepsilon$\\
        Ising model ($d = 3$) & $-0.53\pm0.06$ & Monte Carlo \\
        Ising model ($d = 3$) & $\sim -0.61$ & Padé extrapolation \\
        Gross-Neveu at large $N$ & $ -8.3512\,h^2
    +\mathrm{O}(1/N)$ & Large $N$  \\
        $\mathcal N = 2$ $U(1)_k$ Chern-Simons matter & $ 1.083 k$ & Large $k$ \\
        \hline
    \end{tabular}
    \label{tab:MDS}
\end{table}

The remainder of the paper is organized as follows.
In Section~\ref{sec:ThermalEffectiveTheoryForLineDefects}, we develop the effective theory in Equation~\eqref{eq:LogPartitionFunctionintro} and discuss the physical interpretation of its leading coefficients.
In Section~\ref{sec:asymptotic}, we relate the thermal EFT to asymptotic conformal data and derive Equation~\eqref{eq:AsymptoticRed}.
Finally, in Section~\ref{sec:applications}, we illustrate the EFT in a twisted free scalar theory, compute its leading coefficients for the magnetic line at the $\mathrm{O}(N)$ Wilson--Fisher fixed point, the case of the Gross-Neveu model at large $N$ and the $\mathcal N = 2$ Chern-Simons matter theory and the three-dimensional Ising model via Monte Carlo simulations, comparing the numerical results with perturbative ones.

\section{Thermal effective theory for line defects}
\label{sec:ThermalEffectiveTheoryForLineDefects}

We consider a conformal field theory on $\mathbb{R} \times S_R^{d-1}$ in the presence of a conformal line defect supported on a circle of radius $R\sin\theta$.
The circle lies on a fixed two-sphere $S_R^2\subset S_R^{d-1}$, with $\theta$ denoting its polar angle.
We introduce finite temperature by compactifying the Euclidean time direction on a circle of circumference $\beta=1/T$.
The resulting geometry is $S_\beta^1 \times S_R^{d-1}$, as illustrated in Figure~\ref{fig:finite-temperature-defect-setup}.

In the high-temperature regime, $\beta \ll R$, the nonzero Matsubara modes can be integrated out, yielding an effective field theory in $d-1$ dimensions.
We expect this EFT to simplify further in the presence of thermal screening.
In particular, we assume that connected correlation functions in the presence of the defect decay exponentially at large spatial separations,
\begin{equation}
    \langle \mathcal{O}(x) \mathcal{O}(0) \mathcal{D} \rangle_c
    \sim
    e^{-x/\xi_{\rm scr}}\,,
\end{equation}
where $\mathcal{O}$ is a local bulk operator, $x$ denotes the spatial separation, and $\xi_{\rm scr}$ is the characteristic screening length.
The correlator is understood to be normalized by the defect expectation value.
When the characteristic length scale of the defect is much larger than the screening length, and no additional long-range contributions to the defect free energy are present, we expect the effective theory to be governed by local geometric invariants of the defect embedding.
For the circular defect considered here, this requires
\begin{equation}
    R \sin (\theta)
    \gg
    \xi_{\rm scr}\,.
\end{equation}
For more general curves, we also require that segments well separated along the defect do not approach each other within the screening length.
Under these assumptions, we write down the effective action for a generic smooth spatial defect, independently of its precise shape.

More precisely, on the round spatial sphere, the logarithm of the defect partition function admits the local derivative expansion 
\begin{align}
    \log \left( \frac{Z[\beta,\mathcal{D}]}{Z[\beta]} \right)
    &= \gamma[\mathcal{D}]
    + \int_{\mathcal{D}} \mathrm{d}s \Bigg[
        \frac{m_D}{2\pi\beta}
        + \frac{\beta}{2\pi}
        \left(
            t_2 K_i K^i
            + \frac{t_R}{R^2}
        \right) \notag \\
        &\phantom{=\ }
        + \beta^3 \left(
            t_{4,1}(K_i K^i)^2
            + t_{4,2}(\nabla_s K_i)(\nabla_s K^i)
            + \frac{t_{4,3}}{R^2}K_i K^i
            + \frac{t_{4,4}}{R^4}
        \right)
        + \ldots
    \Bigg]\,.
    \label{eq:LogPartitionFunction}
\end{align}
Here, $Z[\beta,\mathcal{D}]$ and $Z[\beta]$ are the thermal partition functions with and without the defect, and $\mathrm{d}s$ is the proper length element along $\mathcal{D}$.
The vector $K^i$ is the curvature of the line within the spatial sphere, and $\nabla_s$ is the corresponding normal-bundle covariant derivative.
For a latitude $\mathcal{D}_\theta$, the length is $L=2\pi R\sin(\theta)$ and $K_iK^i=\cot^2(\theta)/R^2$.
The coefficients $m_D$, $t_2$, $t_R$, and $t_{4,a}$ are dimensionless and play the following roles:
\begin{itemize}[
  label=$\star$,
  labelindent=0pt,
  labelwidth=1em,
  labelsep=.5em,
  leftmargin=1.5em
]
    \item The term $\gamma[\mathcal{D}]$ is independent of temperature and includes the normalization dependence of the defect.

    \item The coefficient $m_D$ multiplies the line length and is therefore the perimeter coefficient.
    Its leading contribution per unit length is
    \begin{equation}
        \frac{1}{L}
        \left[
            \log \left( \frac{Z[\beta,\mathcal{D}]}{Z[\beta]} \right)
            - \gamma[\mathcal{D}]
        \right]
        =
        \frac{m_D}{2\pi\beta} + \ldots\,.
    \end{equation}
    The corresponding excess free energy per unit length has the opposite sign.

    \item The coefficient $t_2$ measures the response to bending after the perimeter contribution is accounted for, while $t_R$ measures the response to ambient curvature.
    At the equator, the bending term vanishes but the ambient-curvature term remains.

    \item At order $\beta^3$, $t_{4,1}$ and $t_{4,2}$ govern quartic-curvature and curvature-gradient terms, while $t_{4,3}$ and $t_{4,4}$ govern mixed and purely ambient-curvature terms.
\end{itemize}
In the locally flat limit, the terms involving $R^{-2}$ vanish; for a smooth line whose curvature radius is much larger than $\beta$ and the screening length, $t_2$ gives the leading shape correction, while the $\beta^3$ terms are suppressed.
Section~\ref{sec:applications} uses this limit to determine $m_D$ and $t_2$, while Section~\ref{sec:asymptotic} uses $m_D$ to constrain asymptotic conformal data under the assumptions stated there.

\section{Asymptotic conformal data from the thermal EFT}
\label{sec:asymptotic}

In this section, we use the thermal EFT to derive asymptotic constraints on zero-temperature conformal data.
We first express the defect partition function as a thermal trace and determine the asymptotic behavior of defect matrix elements in heavy states.
We then use a convergent bulk OPE representation of primary-state matrix elements to constrain averaged products of defect one-point coefficients and bulk OPE coefficients.

\subsection{High-energy defect matrix elements}
\label{subsec:HighEnergyDefectMatrixElements}

The defect in our setup, shown in Figure~\ref{fig:finite-temperature-defect-setup}, can be inserted into the thermal trace over the Hilbert space on the spatial sphere:
\begin{equation}
    Z[\beta,\mathcal{D}_\theta]
    = \sum_n e^{-\beta\Delta_n/R}
      \langle n|\mathcal{D}_\theta|n\rangle\,,
      \label{eq:Z1}
\end{equation}
where $\Delta_n$ denotes the scaling dimension of the state $|n\rangle$, and the sum includes both primary states and their descendants.
To study the high-energy asymptotics, it is convenient to express this trace in terms of the density of all states:
\begin{equation}
    Z[\beta,\mathcal{D}_\theta]
    = \int \mathrm{d}\Delta\,\rho_{\mathrm{all}}(\Delta)\,
      \overline{\langle\Delta|\mathcal{D}_\theta|\Delta\rangle}_{\mathrm{all}}\,
      e^{-\beta\Delta/R}\,.
    \label{eq:Z}
\end{equation}
Here, $\rho_{\mathrm{all}}(\Delta)=e^{S_{\mathrm{all}}(\Delta)}$ is the density of states, and the overline denotes an average over all states with scaling dimension near $\Delta$ including both primaries and descendants.
The thermal trace converges for $\beta>0$, while the thermal EFT derived in Section~\ref{sec:ThermalEffectiveTheoryForLineDefects} applies at high temperature, provided the defect is much larger than the screening length.
In this regime, it predicts
\begin{equation}
    \log \left( Z[\beta,\mathcal{D}_\theta] \right)
    = -\frac{fR^{d-1}}{\beta^{d-1}}
      +\gamma+\frac{m_D R}{\beta}\sin(\theta)+\ldots\,,
    \label{eq:highT}
\end{equation}
where $f<0$ is the bulk free-energy coefficient.

For comparison, the partition function without the defect is
\begin{equation}
    Z[\beta]
    = \int \mathrm{d}\Delta\,\rho_{\mathrm{all}}(\Delta)e^{-\beta\Delta/R}\,,
    \label{eq:bulkZ}
\end{equation}
with the high-temperature expansion
\begin{equation}
    \log \left( Z[\beta] \right)
    = -\frac{fR^{d-1}}{\beta^{d-1}}+\ldots\,.
    \label{eq:bulkEFT}
\end{equation}
The leading term determines the leading heavy-state entropy; subleading corrections and further applications are discussed in~\cite{Benjamin:2023qsc}.
We denote by $\gamma_{\mathrm{rel}}$ the constant term in $\log[Z[\beta,\mathcal D_\theta]/Z[\beta]]$; it equals $\gamma$ if no constant bulk term has been omitted from~\eqref{eq:bulkEFT}.

To extract data of \emph{primary} states, we must first organize the trace into conformal multiplets.
Let $q=e^{-\beta/R}$ and let $p$ labels a primary of dimension $\Delta_p$ and spin $J_p$.
The exact defect trace can be written as
\begin{equation}
    Z[\beta,\mathcal D_\theta]
    =\sum_p q^{\Delta_p}\mathcal F^{\mathcal D}_{p}(q,\theta),
    \qquad
    \mathcal F^{\mathcal D}_{p}(q,\theta)
    :=q^{-\Delta_p}\operatorname{Tr}_{p}
       \bigl(q^D\mathcal D_\theta\bigr) \ ,
    \label{eq:defect-thermal-block}
\end{equation}
where $\operatorname{Tr}_{p}$ operates on the conformal multiplet associated to the primary operator labeled by $p$: the defect thermal block $\mathcal F^{\mathcal D}_{p}$ includes every descendant of $p$.
For a generic long multiplet, the corresponding unrefined character is $q^{\Delta_p}\dim(J_p)/(1-q)^d$.
Define $M_p(\theta)$ as the trace of $\mathcal D_\theta$ over the primary spin multiplet, divided by $\dim(J_p)$.
For the heavy multiplets contributing at the inverse-Laplace saddle, the defect thermal block is expected to obey
\begin{equation}
    \mathcal F^{\mathcal D}_{p}(q,\theta)
    \simeq \frac{\dim(J_p)}{(1-q)^d}M_p(\theta).
    \label{eq:defect-block-heavy}
\end{equation}
This holds uniformly as $q\to1$ in the angular regime used below.
Writing 
\[\mathcal P_0(q)=\sum_p\dim(J_p)q^{\Delta_p} \ ,\hspace{0.5 cm} \ \mathcal P_D(q,\theta)=\sum_p\dim(J_p)M_p(\theta)q^{\Delta_p} \ ,\] we then have
\begin{equation}
    Z[\beta]\simeq(1-q)^{-d}\mathcal P_0(q),
    \qquad
    Z[\beta,\mathcal D_\theta]\simeq(1-q)^{-d}\mathcal P_D(q,\theta).
    \label{eq:primary-generating-functions}
\end{equation}

An inverse Laplace transform of~\eqref{eq:bulkEFT}, evaluated by the saddle-point method, gives the Cardy-type asymptotics
\begin{align}
    \frac{\beta_*}{R}
    &= \left(\frac{-(d-1)f}{\Delta}\right)^{1/d},
    \label{eq:saddle}\\
    S_{\mathrm{all}}(\Delta)
    &= \frac{d}{d-1}\bigl(-(d-1)f\bigr)^{1/d}
       \Delta^{(d-1)/d}.
    \label{eq:DOS}
\end{align}
As $\Delta\to\infty$, the saddle approaches the high-temperature limit, $\beta_*/R\to0$.
The primary density has the same leading exponential growth: for long multiplets, $\rho_{\mathrm{prim}}(\Delta)\sim(\beta_*/R)^d\rho_{\mathrm{all}}(\Delta)$, so removing descendants changes only the power-law prefactor.

More precisely, define the primary density and the defect-weighted primary density by
\begin{align}
    \rho_{\mathrm{prim}}(\Delta)
    &=\sum_p\dim(J_p)\delta(\Delta-\Delta_p),\\
    \rho_{\mathrm{prim}}(\Delta)\,
    \overline{M}_{\mathrm{prim}}(\Delta,\theta)
    &=\sum_p\dim(J_p)M_p(\theta)\delta(\Delta-\Delta_p).
    \label{eq:primary-densities}
\end{align}
By~\eqref{eq:primary-generating-functions}, these densities are obtained by inverse Laplace transforming $(1-q)^d Z[\beta]$ and $(1-q)^d Z[\beta,\mathcal D_\theta]$, respectively.
For $d>2$, the defect term is subleading to the bulk saddle-point action, and the common character factor cancels in their ratio.
In the small-angle regime $\beta_*/R\ll\sin\theta\ll1$ used below, this gives
\begin{equation}
    \overline{M}_{\mathrm{prim}}(\Delta,\theta)
    \simeq e^{\gamma_{\mathrm{rel}}+A(\Delta)\sin\theta},
    \label{eq:ETHdefect}
\end{equation}
where
\begin{equation}
    A(\Delta)=\frac{m_D R}{\beta_*}
    =m_D\left(\frac{\Delta}{-(d-1)f}\right)^{1/d}.
    \label{eq:A}
\end{equation}
At fixed $\theta$, the leading term $A(\Delta)\sin\theta$ still controls the logarithm, but in $d=3$ the shift of the thermal saddle can affect the order-one prefactor.

\subsection{Bulk OPE and asymptotic conformal data}
\label{subsec:BulkOPEAndAsymptoticConformalData}

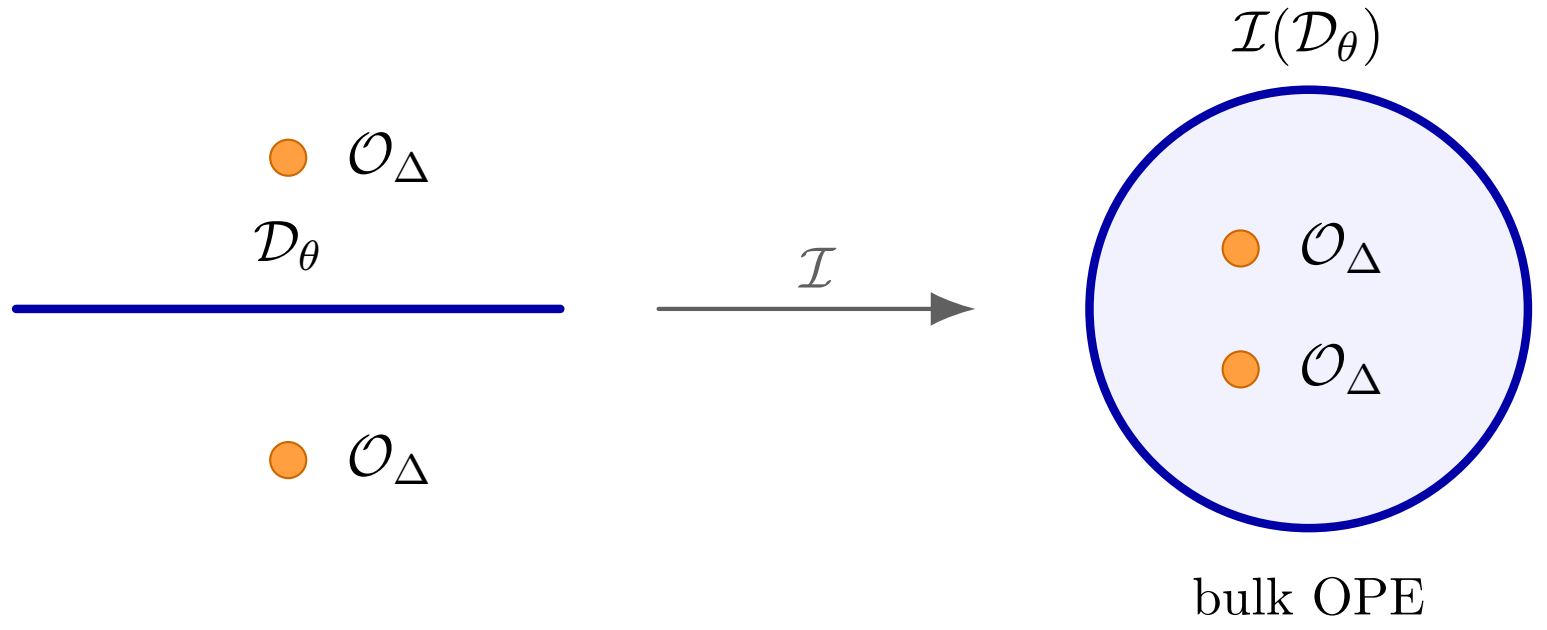
\begin{figure}[t]
    \centering
    \begin{tikzpicture}[
        line cap=round,
        defect/.style={draw=blue!65!black, line width=1.6pt},
        insertion/.style={
            circle,
            draw=orange!80!black,
            fill=orange!75,
            inner sep=2.4pt
        }
    ]
        \node[insertion] at (-3.5,1.0) {};
        \node[anchor=west] at (-3.25,1.0) {$\mathcal{O}_\Delta$};

        \draw[defect] (-5.3,0) -- (-1.7,0);
        \node[above] at (-3.5,0.12) {$\mathcal{D}_\theta$};

        \node[insertion] at (-3.5,-1.0) {};
        \node[anchor=west] at (-3.25,-1.0) {$\mathcal{O}_\Delta$};

        \draw[-{Latex[length=3mm]}, thick, gray!75!black]
            (-1.05,0) -- (1.05,0)
            node[midway,above] {$\mathcal{I}$};

        \draw[defect, fill=blue!5] (3.25,0) circle (1.45);
        \node at (3.25,1.8) {$\mathcal{I}(\mathcal{D}_\theta)$};

        \node[insertion] at (2.8,0.4) {};
        \node[anchor=west] at (3.05,0.4) {$\mathcal{O}_\Delta$};

        \node[insertion] at (2.8,-0.4) {};
        \node[anchor=west] at (3.05,-0.4) {$\mathcal{O}_\Delta$};

        \node[font=\small] at (3.25,-1.9) {bulk OPE};
    \end{tikzpicture}
    \caption{A conformal inversion maps the straight line defect to a circle.
    In the transformed configuration, the two bulk insertions of a scalar primary lie in a region where the bulk OPE converges.}
    \label{fig:inversion-bulk-ope}
\end{figure}

We now apply the bulk OPE to the matrix element of an individual \emph{primary}, before averaging over primary states.
For the explicit scalar OPE below, we take the heavy operator $\mathcal O_p$ to be a scalar primary of dimension $\Delta_p\simeq\Delta$.
The unrefined trace in~\eqref{eq:Z1} also contains spinning primaries; using~\eqref{eq:ETHdefect} for the scalar-primary average therefore requires a scalar-spin projection whose defect-to-bulk ratio has the same $m_D\sin\theta/t$ term and, for the prefactor below, the same $\gamma_{\mathrm{rel}}$, or an independent treatment of spinning primaries.
We make this additional assumption in the following matching.
At first sight, the matrix element $\langle p|\mathcal{D}_\theta|p\rangle$ does not appear to admit a bulk OPE expansion.
In the corresponding flat-space configuration, the defect lies between the two insertions of $\mathcal O_p$.
However, a suitable conformal inversion maps the straight defect to a circle, leaving the insertions in a common region where the bulk OPE converges, as illustrated in Figure~\ref{fig:inversion-bulk-ope}.
The resulting bulk OPE involves one-point functions of local operators in the presence of the defect, whose form is fixed by the preserved conformal symmetry up to their coefficients.
Normalizing by $g_D=\langle0|\mathcal D_\theta|0\rangle$ and denoting the normalized defect one-point coefficient by $a_\chi$, we obtain
\begin{equation}
    \frac{\langle p|\mathcal{D}_\theta|p\rangle}{g_D}
    =\sum_\chi a_\chi f_{\mathcal O_p\mathcal O_p\chi}
      2^{\Delta_\chi}\mathcal B_\chi(\theta)\,.
    \label{eq:defOPe}
\end{equation}
Here $f_{\mathcal O_p\mathcal O_p\chi}$ is the bulk OPE coefficient and $\mathcal B_\chi(\theta)$ is the corresponding conformal block.
Although these blocks are not known in closed form in general dimensions, their large-$\Delta_\chi$ behavior is known~\cite{Lauria:2017wav,Lauria:2018klo}:
\begin{equation}
    \mathcal B_\chi(\theta)
    \sim \mathcal T_{J_\chi}r^{\Delta_\chi}
    (1-r^2)^{\frac{3-d}{2}}(1+r^2)^{-\frac32}\,.
    \label{eq:defect-bulk-block-asymptotic}
\end{equation}
Here $r=\tan(\theta/2)$, while $\mathcal T_{J_\chi}=\mathfrak C_{J_\chi}(0)$ encodes the spin dependence, with $\mathfrak C_{J_\chi}$ denoting the angular polynomials for the exchanged symmetric traceless representations~\cite{Lauria:2017wav,Lauria:2018klo}.
For identical scalar external operators, only even spins contribute.
The asymptotic form~\eqref{eq:defect-bulk-block-asymptotic} must also remain valid in the double limit used below, where $\Delta_\chi$ becomes large and $r$ becomes small.

The descendants of $p$ are already included in the defect thermal block~\eqref{eq:defect-thermal-block}; this is why the thermal inversion can be interpreted as a statement about primaries only after~\eqref{eq:defect-block-heavy} is imposed.
Let $\overline{f}_{\Delta,\chi}^{\mathrm{prim},0}$ denote the average of $f_{\mathcal O_p\mathcal O_p\chi}$ over scalar primaries with $\Delta_p$ in a narrow window around $\Delta$, and set
\begin{equation}
    c_{\Delta,\chi}:=a_\chi\mathcal T_{J_\chi}
    \overline{f}_{\Delta,\chi}^{\mathrm{prim},0}.
    \label{eq:average-ope-coefficient}
\end{equation}
Using the assumed scalar-primary version of~\eqref{eq:ETHdefect} and retaining the leading term at small $r$, we obtain
\begin{equation}
    \sum_\chi c_{\Delta,\chi}(2r)^{\Delta_\chi}
    \simeq e^{\widehat\gamma+A\sin\theta},
    \qquad
    \widehat\gamma=\gamma_{\mathrm{rel}}-\log g_D\,,
    \label{eq:ope-thermal-matching}
\end{equation}
where $A=A(\Delta)$ was defined in~\eqref{eq:A}.
The inverse transform in the angular variable constrains the OPE-weighted spectral density $c_\Delta(\delta):=\sum_\chi c_{\Delta,\chi}\delta(\delta-\Delta_\chi)$, understood after suitable smoothing in $\delta$; it does not isolate an individual OPE coefficient.

To extract the contribution at large dimension $\Delta_\chi$, we formally invert this expansion and evaluate the resulting integral by the saddle-point method.
The saddle-point condition is 
\begin{equation}
    \sin(2\theta_*)=\frac{2\Delta_\chi}{A}\,.
    \label{eq:ope-saddle}
\end{equation}
For $A>0$, the relevant real saddle lies on the branch $\theta_*<\pi/4$.
In the regime $\Delta_\chi\ll A$, it approaches
\begin{equation}
    \theta_*\simeq\frac{\Delta_\chi}{A}\,.
\end{equation}
The saddle-point approximation then gives
\begin{equation}
    \begin{aligned}
    c_\Delta(\Delta_\chi)
    &\simeq
    \frac{e^{\widehat\gamma}}{\sqrt{2\pi\Delta_\chi}}
    \left(\frac{eA}{\Delta_\chi}\right)^{\Delta_\chi}
    \simeq e^{\widehat\gamma}
    \frac{A^{\Delta_\chi}}{\Gamma(\Delta_\chi+1)}\\
    &=\frac{e^{\widehat\gamma}m_D^{\Delta_\chi}}
    {\Gamma(\Delta_\chi+1)}
    \left(\frac{\Delta}{-(d-1)f}\right)^{\Delta_\chi/d}\,,
    \end{aligned}
    \label{eq:ope-asymptotic-2r}
\end{equation}
where the asymptotic relation concerns a suitably smoothed density in $\Delta_\chi$ and assumes that cancellations among signed OPE weights do not invalidate the saddle estimate.
The saddle lies within the regime of validity of the thermal EFT provided
\begin{equation}
    \max(1,m_D)\ll\Delta_\chi\ll A\,.
\end{equation}
Assuming that the EFT coefficients are of order $\mathrm O(1)$,\footnote{This requirement excludes certain large-$N$ limits.} the parametric window is $1\ll\Delta_\chi\ll\Delta^{1/d}$.
The result~\eqref{eq:ope-asymptotic-2r} is a constraint on the averaged spectral density $c_\Delta(\Delta_\chi)$, with $\mathcal T_{J_\chi}$ included in its definition~\eqref{eq:average-ope-coefficient}.

Note that what done before is rigorously speaking only valid for $A>0$ which means for $m_D>0$. In the example below we will show that for some models the value of $m_D$ is positive while in other cases, e.g. the Ising model, is negative. In the cases in which $m_D$ is negative, the derivation of the asymptotic is less trivial: a direct analytic continuation of equation \eqref{eq:ope-asymptotic-2r} would bring a complex phase of the form $(-1)^{\Delta_\chi}$ \footnote{In principle we can study $\tilde{c}_{\Delta,\chi} = (-1)^{\Delta_\chi} \mathcal{T}_{J_\chi} \bar{f}_{\Delta,\chi}^{\rm prim, 0}$ }: this is clearly not physical and has to be interpreted in an average sense. A less disturbing way of presenting this result is to consider the absolute value in both sided: mathematically this is justified with the additional assumption that the thermal EFT proposed in this paper rule the partition function for $|\beta|\to \infty$ in the whole complex $\beta$-plane.  

Before concluding, we observe that another possible expansion expresses the defect itself in terms of local bulk operators~\cite{Berenstein:1998ij}.
This expansion applies when the defect is sufficiently small and has a finite radius of convergence, as can be seen in simple free-field examples at finite temperature.
In our setup, the relevant small-defect regime is $R\sin\theta\ll\beta$, in contrast with the large-defect regime required by the thermal EFT.
The two expansions therefore describe different limits of the defect partition function, making a direct comparison between them less straightforward.

\section{Applications}
\label{sec:applications}

In this section, we examine the scope and coefficients of the thermal line EFT in three examples.
For a free scalar, periodic thermal boundary conditions leave an unscreened mode, while a $\mathbb Z_2$-twisted trace gives a screened example whose perimeter and shape coefficients can be computed exactly.
At the $\mathrm{O}(N)$ Wilson--Fisher fixed point, we compute the magnetic line's perimeter coefficient $m_D$ through order $\sqrt{\veps}$ and its leading bending coefficient $t_2$.
Finally, Monte Carlo simulations of the critical three-dimensional Ising model provide an estimate of $m_D$, which we compare with a Padé interpolation of the $N=1$ result constrained by the exact $d=2$ value.

\subsection{Free scalar field}
\label{subsec:FreeScalar}

We begin with the simplest example: a line defect in the free scalar theory in $d=4$.
Since the theory is Gaussian, the defect expectation value is computed exactly by a single propagator and the flat-space coefficients can be obtained in closed form.
Unfortunately, with the standard thermal boundary conditions the defect is not screened, and the EFT does not apply.
Screening, and with it the EFT, is restored by twisting the boundary conditions along the thermal circle by the $\mathbb{Z}_2$ symmetry $\phi\to-\phi$.

\subsubsection{Setup}
\label{subsubsec:FreeSetup}

\paragraph{Action.}
The bulk action and line insertion are given by
\begin{align}
    S_{\mathrm{bulk}}
    &=
    \int \mathrm{d}^4 x\,\sqrt{g}
    \left[
        \frac{1}{2}(\nabla\phi)^2
        +
        \frac{\xi_4}{2}\mathcal R\,\phi^2
    \right],
    \qquad
    \xi_4=\frac{1}{6},
    \\
    D(C)
    &=
    \exp\!\left[
        -h
        \int_C\mathrm{d}\ell\, \phi
    \right].
\end{align}
Since $\Delta_\phi=1$, the coupling $h$ is exactly marginal and $D(C)$ is a conformal line defect for every value of $h$.
Because the theory is Gaussian, the logarithm of the defect expectation value is given exactly by a single propagator joining two source insertions,
\begin{equation}
    W_\beta[C]
    =
    \log \langle D(C)\rangle_\beta
    =
    \frac{h^2}{2}
    \int_C\mathrm{d}\ell
    \int_C\mathrm{d}\ell'\,
    G_\beta\bigl(x(\ell),x(\ell')\bigr)
    =
    \begin{tikzpicture}[scale = 1, baseline={([yshift=-.8ex]current bounding box.center)}]
        \draw[wilson] (-.25,0) -- (1.75,0);
        \draw[scalar] (.25,0) arc (180:0:.5);
        \draw[VertexColor,fill=VertexColor] (.25,0) circle (.075);
        \draw[VertexColor,fill=VertexColor] (1.25,0) circle (.075);
    \end{tikzpicture}
    \,.
    \label{eq:FreeExchange}
\end{equation}
We use dimensional regularization with temperature-independent counterterms, as in the rest of this section.
In particular, the power divergence at coincident points does not contribute.

\paragraph{Boundary conditions.}
We consider two choices of boundary conditions along the thermal circle,
\begin{equation}
    \phi(\tau+\beta,\bm{x})
    =
    \pm\,\phi(\tau,\bm{x}).
    \label{eq:FreeBoundaryConditions}
\end{equation}
The periodic choice computes the standard thermal trace.
The antiperiodic choice inserts the $\mathbb{Z}_2$ generator $g:\phi\to-\phi$ in the trace, and computes
\begin{equation}
    W^-_\beta[C]
    =
    \log
    \frac{\operatorname{Tr}\bigl(g\,D(C)\,e^{-\beta H}\bigr)}
         {\operatorname{Tr}\bigl(g\,e^{-\beta H}\bigr)}.
\end{equation}
We denote the periodic result by $W^+_\beta[C]$.
The defect is odd under $g$, so moving the twist across the line flips $h\to-h$.
Since $W^\pm_\beta[C]$ depends only on $h^2$, the position of the twist is immaterial.
Now we will consider a circle $C_\rho$ of radius $\rho$ in flat space.
The integrals along the circle can be performed in closed form, giving
\begin{equation}
    W^\pm_\beta[C_\rho]
    =
    h^2\rho^2\sum_{n=1}^{\infty}
    \frac{(\pm1)^n}{n\beta\sqrt{n^2\beta^2+4\rho^2}}.
    \label{eq:FreeFlatCircle}
\end{equation}

\subsubsection{Periodic scalar}
\label{subsubsec:PeriodicScalar}

For periodic boundary conditions, the sum is dominated by $n\lesssim\rho/\beta$, and
\begin{equation}
    W^+_\beta[C_\rho]
    =
    \frac{h^2\rho}{2\beta}
    \left[
        \log\!\left(\frac{4\rho}{\beta}\right)
        +\gamma_{\mathrm E}
    \right]
    +
    \mathrm{O}\!\left(\frac{\beta}{\rho}\right)\,.
    \label{eq:FreeFlatCirclePeriodic}
\end{equation}
This logarithmic growth lies outside the local thermal EFT because the periodic static mode is unscreened.
But it is not surprising since for periodic boundary conditions we expect an appearance of the gapless mode.
The same computation applies to the setup of Section~\ref{sec:asymptotic}, with the latitude $\mathcal D_\theta$ on $S^1_\beta\times S^3_R$.
Using the propagator on $\mathbb{R}\times S^3_R$, we find
\begin{equation}
    W^\pm_\beta[\mathcal D_\theta]
    =
    \frac{h^2\sin^2\theta}{4}
    \sum_{n=1}^{\infty}
    \frac{(\pm1)^n}
    {\sinh\!\left(\frac{n\beta}{2R}\right)
     \sqrt{\sinh^2\!\left(\frac{n\beta}{2R}\right)+\sin^2\theta}}.
    \label{eq:FreeSphereLatitude}
\end{equation}
On the sphere, the conformal coupling gives the static mode a mass $1/R$, so the periodic sum converges.
The logarithm is then cut off by the radius of the sphere,
\begin{equation}
    \log\frac{Z[\beta,\mathcal D_\theta]}{Z[\beta]}
    =
    \frac{h^2R\sin(\theta)}{2\beta}
    \left[
        \log\!\left(\frac{R}{\beta}\right)
        +
        \log\!\left(\frac{4\sin(\theta)}{1+\sin(\theta)}\right)
        +\gamma_{\mathrm E}
    \right]
    +
    \mathrm{O}\!\left(\frac{\beta}{R}\right).
    \label{eq:FreeSpherePeriodic}
\end{equation}
In the limit $\theta\to0$ at fixed $R\sin(\theta)$, this reproduces~\eqref{eq:FreeFlatCirclePeriodic}.

\subsubsection{Antiperiodic scalar}
\label{subsubsec:AntiperiodicScalar}

With antiperiodic boundary conditions, the images in~\eqref{eq:FreeSphereLatitude} alternate in sign, and the harmonic series is replaced by a convergent alternating series,
\begin{equation}
    \frac{W^-_\beta[C]}{L}
    =
    \frac{h^2}{4\pi\beta}\sum_{n=1}^{\infty}\frac{(-1)^n}{n}
    =
    -\frac{h^2\log 2}{4\pi\beta}.
\end{equation}
Matching to~\eqref{eq:StraightLinePerimeterMatching}, we obtain
\begin{equation}
    m_D=-\frac{h^2}{2}\log 2.
    \label{eq:FreeTwistedmD}
\end{equation}
For antiperiodic boundary conditions we can even make more.
Since $G^-_\beta$ decays exponentially, only points separated by $|\ell-\ell'|\lesssim\beta$ contribute to~\eqref{eq:FreeExchange} for a smooth curve.
The full derivative expansion of~\eqref{eq:LogPartitionFunction} can then be obtained by expanding the distance between two points of the curve.
Let $s$ be the proper length along $C$, $u=s'-s$, $T^i=\dot x^i$ the unit tangent vector, and $K^i=\ddot x^i$ the extrinsic-curvature vector.
Then
\begin{equation}
    |\bm{x}(s+u)-\bm{x}(s)|^2
    =
    u^2
    -\frac{K_iK^i}{12}u^4
    -\frac{K_i\dot K^i}{12}u^5
    -\left(
        \frac{\dot K_i\dot K^i}{45}
        +\frac{K_i\ddot K^i}{40}
    \right)u^6
    +\mathrm{O}(u^7),
\end{equation}
where dots denote derivatives with respect to $s$ in the ambient space.
We insert this into~\eqref{eq:FreeExchange} and use $\dot K^i=\nabla_sK^i-(K_jK^j)T^i$.
After integrating by parts along $C$, this gives
\begin{equation}
    W^-_\beta[C]
    =
    h^2\int_C\mathrm{d}s
    \left[
        M_0
        +\frac{M_2}{8}K_iK^i
        +M_4\left(
            \frac{3}{128}(K_iK^i)^2
            -\frac{1}{144}(\nabla_sK_i)(\nabla_sK^i)
        \right)
        +\ldots
    \right],
    \label{eq:FreeTwistedLocal}
\end{equation}
where the moments of the propagator are
\begin{equation}
    M_{2k}
    \equiv
    \int_0^\infty\mathrm{d}v\,v^{2k}\,G^-_\beta(v)
    =
    (-1)^{k+1}\,\frac{\eta(1-2k)}{4\pi}\,\beta^{2k-1},
    \label{eq:FreeTwistedMoments}
\end{equation}
and $\eta(s)=\sum_{n\geq1}(-1)^{n-1}n^{-s}$ is the Dirichlet eta function.
For $k=0$, the integral is defined in dimensional regularization.
Explicitly, $M_0=-\log2/(4\pi\beta)$, $M_2=\beta/(16\pi)$ and $M_4=\beta^3/(32\pi)$.

Matching~\eqref{eq:FreeTwistedLocal} to~\eqref{eq:LogPartitionFunction}, we obtain
\begin{equation}
    \begin{aligned}
        m_D&=-\frac{h^2}{2}\log2,
        &\qquad
        t_2&=\frac{h^2}{64},
        \\
        t_{4,1}&=\frac{3h^2}{4096\pi},
        &\qquad
        t_{4,2}&=-\frac{h^2}{4608\pi}.
    \end{aligned}
    \label{eq:FreeTwistedCoefficients}
\end{equation}
These results are exact in $h$.
Every term in~\eqref{eq:FreeTwistedLocal} is an odd power of $\beta$, so $\gamma[\mathcal D]=0$ up to corrections exponentially small in the size of the defect.
The bending coefficient $t_2$ can also be computed from the general Gaussian formula $t_2=\tfrac{\pi h^2}{8\beta}\int_{-\infty}^{\infty}\mathrm{d}v\,v^2G(v)$.

In the language of Section~\ref{sec:asymptotic}, the twisted trace weighs each state by its $\mathbb{Z}_2$ charge, $Z_g[\beta,\mathcal D_\theta]=\sum_n g_n\,e^{-\beta\Delta_n/R}\langle n|\mathcal D_\theta|n\rangle$.
In the interacting examples below, no twist is needed: screening is generated dynamically by the thermal mass.

\subsection{Magnetic line at the Wilson--Fisher fixed point}
\label{subsec:MagneticLineAtTheWFFixedPoint}

We consider the $\mathrm{O}(N)$ Wilson--Fisher theory in $d=4-\veps$ on $S^1_\beta \times S_R^{d-1}$, with a magnetic line embedded in the spatial sphere at fixed Euclidean time.
The goal is to compare explicit calculations in the regime $\veps\ll1$ to the effective action~\eqref{eq:LogPartitionFunction} in order to determine $m_D$ and $t_2$.

\subsubsection{Setup}
\label{subsubsec:ActionAndConventions}

\paragraph{Action.}
The bulk action and line insertion are given by
\begin{align}
    S_{\mathrm{bulk}}
    &=
    \int \mathrm{d}^d x\,\sqrt{g}
    \left[
        \frac{1}{2}(\nabla\phi^I)^2
        +
        \frac{\xi_d}{2}\mathcal R\,\phi^I\phi^I
        +
        \frac{\mu^\veps\lambda}{4!}(\phi^I\phi^I)^2
    \right],
    \\
    D(C)
    &=
    \exp\!\left[
        -\mu^{\veps/2} h
        \int_C\mathrm{d}\ell\, \phi^1
    \right],
\end{align}
with $I=1,\ldots,N$.
Here, $\xi_d$ is the conformal curvature coupling, $g=\det(g_{\mu\nu})$ is the determinant of the metric, and $\mathcal{R}$ is the scalar curvature of $S^1_\beta \times S_R^{d-1}$, while $\mathrm{d}\ell$ is the proper length element along $C$.
We set the renormalization scale to $\mu=1/\beta$ and define
\begin{equation}
    W_\beta[C]
    =
    \log\left(\langle D(C)\rangle_\beta\right)
    =
    \log\left(
        \frac{Z[\beta,D(C)]}{Z[\beta]}
    \right).
\end{equation}
Note that counterterms are temperature-independent and thus we can use the zero-temperature ones.

\paragraph{Fixed point and thermal mass.}
At the Wilson--Fisher magnetic-line fixed point, the couplings in our normalization obey
\begin{align}
    \lambda_\star
    &=
    \frac{48\pi^2}{N+8}\veps+\mathrm{O}(\veps^2),
    \\
    h_\star^2
    &=
    N+8+\mathrm{O}(\veps).
\end{align}
Since $\lambda_\star=\mathrm{O}(\veps)$, the thermal theory has separated scales.
The zero-frequency mode acquires a mass of order $\sqrt{\veps}/\beta$.
Although the line at fixed Euclidean time sources all modes, its response at distances much larger than $\beta$ is governed by the zero mode.
We define the static mode by
\begin{equation}
    \phi_0^I(\bm{x})
    =
    \frac{1}{\beta}
    \int_0^\beta\mathrm{d}\tau\,\phi^I(\tau,\bm{x}).
\end{equation}
Integrating out the nonzero modes gives an effective $(d-1)$-dimensional theory for $\phi_0^I(\bm{x})$.
Its line-sourced profile supplies the long-distance part of the perimeter coefficient in~\eqref{eq:LogPartitionFunction}.
The thermal mass of the bulk theory can be written at leading order as
\begin{equation}
    m_0^2
    =
    \frac{N+2}{72\beta^2}\lambda_\star.
\end{equation}
Here $m_0$ is the leading mass parameter of the reduced theory; fluctuations of the static field correct the physical screening mass at higher order.

To extract the perimeter coefficient of~\eqref{eq:LogPartitionFunction}, we consider a portion of the line whose curvature radius is much larger than the screening length $m_0^{-1}$.
For a latitude at angle $\theta$ on the spatial sphere, this requires $m_0 R \sin(\theta) \gg 1$.
Imposing this condition before taking $\veps\to0$ at fixed $N$, this gives the hierarchy
\begin{equation}
    \beta \ll m_0^{-1} \ll R \sin (\theta).
\end{equation}
The response on the screening scale can therefore be computed using a \textit{straight line in flat space}.

\subsubsection{The perimeter coefficient $m_D$}
\label{subsubsec:PerimeterCoefficient}

For an infinite straight line, the perimeter coefficient is determined by the term in $W_\beta[C]$ proportional to the line length,
\begin{equation}
    \lim_{L\to\infty}\frac{W_\beta[C]}{L}
    =
    \frac{m_D}{2\pi\beta},
    \label{eq:StraightLinePerimeterMatching}
\end{equation}
where $L$ denotes the length of the line.
We determine both the logarithmic and finite terms as $\veps\to0$ at fixed $N$.

Since $W_\beta[C]$ is the logarithm of the defect expectation value, it is computed by connected diagrams with source insertions on the line.
The simplest contribution joins two source insertions by a single bulk propagator.
We first evaluate this Gaussian contribution, with the leading thermal mass included in the static propagator.
We then include diagrams with quartic interactions between static fields.
As we explain below, these also contribute to the finite term and are summed by a nonlinear static profile.

\paragraph{Gaussian contribution.}
The Gaussian contribution comes from a single bulk propagator joining two source insertions on the line,
\begin{equation}
    \begin{tikzpicture}[scale = 1, baseline={([yshift=-.8ex]current bounding box.center)}]
        \draw[wilson] (-.25,0) -- (1.75,0);
        \draw[scalar] (.25,0) arc (180:0:.5);
        \draw[VertexColor,fill=VertexColor] (.25,0) circle (.075);
        \draw[VertexColor,fill=VertexColor] (1.25,0) circle (.075);
    \end{tikzpicture}
    \,.
    \label{eq:Gaussian_Diagram}
\end{equation}
The propagator includes both the nonzero Matsubara modes and the static mode with its leading thermal mass.
This accounts for distances of order $\beta$, where the reduced static theory alone is insufficient.
The propagator is evaluated in the bulk thermal theory without the defect insertion; the defect enters through the two source vertices.

Let $v=\ell-\ell'$ denote the separation between two points on the straight line at the same Euclidean time.
We denote by $G(|v|)$ the propagator of $\phi^1$ between these points, with the leading thermal mass included in its static mode.
At the order needed here, we evaluate it in four dimensions.
For $v>0$, it is given by
\begin{equation}
    G(v)
    =
    \frac{\coth(\pi v/\beta)}{4\pi\beta v}
    +
    \frac{e^{-m_0 v}-1}{4\pi\beta v}.
    \label{eq:ResummedLinePropagator}
\end{equation}
The first term is the massless thermal propagator.
The second replaces its massless static contribution by the massive one.

The Gaussian contribution to $W_\beta[C]$, denoted by $W_\beta^{(2)}$, is quadratic in the line source:
\begin{equation}
    \frac{W_\beta^{(2)}}{L}
    =
    \frac{h_\star^2}{2 L}
    \int_C\mathrm{d}\ell
    \int_C\mathrm{d}\ell'\,
    G(|\ell-\ell'|)
    =
    -\frac{h_\star^2}{4\pi\beta}\log(\beta m_0).
\end{equation}
where we regulated coincident points with a small separation cutoff.
This gives the logarithmic contribution to the perimeter coefficient, with its finite normalization fixed by including all Matsubara modes.
It also includes the Gaussian static contribution, which must be accounted for when adding the nonlinear correction.

\paragraph{Nonlinear static contribution.}
We now compute the finite correction from connected static tree diagrams with quartic bulk vertices.
We sum these diagrams by evaluating the nonlinear saddle and subtracting the Gaussian static contribution already included above:
\begin{equation}
    \begin{tikzpicture}[scale=0.85, baseline={([yshift=-.8ex]current bounding box.center)}]
        \draw[wilson] (-.15,0) -- (1.75,0);
        \foreach \x in {0,.5,1.1,1.6} {
            \draw[scalar] (\x,0) -- (.8,1);
            \draw[VertexColor,fill=VertexColor] (\x,0) circle (.065);
        }
        \draw[VertexColor,fill=VertexColor] (.735,.935) rectangle (.865,1.065);
    \end{tikzpicture}
    \;+\;
    \begin{tikzpicture}[scale=0.85, baseline={([yshift=-.8ex]current bounding box.center)}]
        \draw[wilson] (-.15,0) -- (2.35,0);
        \draw[scalar] (.35,1) -- (1.85,1);
        \foreach \x in {0,.35,.7} {
            \draw[scalar] (\x,0) -- (.35,1);
            \draw[VertexColor,fill=VertexColor] (\x,0) circle (.065);
        }
        \foreach \x in {1.5,1.85,2.2} {
            \draw[scalar] (\x,0) -- (1.85,1);
            \draw[VertexColor,fill=VertexColor] (\x,0) circle (.065);
        }
        \foreach \x in {.35,1.85} {
            \draw[VertexColor,fill=VertexColor] ({\x-.065},.935) rectangle ({\x+.065},1.065);
        }
    \end{tikzpicture}
    \;+\;
    \begin{tikzpicture}[scale=0.85, baseline={([yshift=-.8ex]current bounding box.center)}]
        \draw[wilson] (-.15,0) -- (3.15,0);
        \draw[scalar] (.4,1) -- (1.5,1) -- (2.6,1);
        \foreach \x in {0,.4,.8} {
            \draw[scalar] (\x,0) -- (.4,1);
            \draw[VertexColor,fill=VertexColor] (\x,0) circle (.065);
        }
        \foreach \x in {1.3,1.7} {
            \draw[scalar] (\x,0) -- (1.5,1);
            \draw[VertexColor,fill=VertexColor] (\x,0) circle (.065);
        }
        \foreach \x in {2.2,2.6,3} {
            \draw[scalar] (\x,0) -- (2.6,1);
            \draw[VertexColor,fill=VertexColor] (\x,0) circle (.065);
        }
        \foreach \x in {.4,1.5,2.6} {
            \draw[VertexColor,fill=VertexColor] ({\x-.065},.935) rectangle ({\x+.065},1.065);
        }
    \end{tikzpicture}
    \;+\;\cdots\,.
    \label{eq:StaticTreeDiagrams}
\end{equation}
Circles denote source insertions, squares represent quartic vertices, and thin lines are static propagators with thermal mass $m_0$; the thick horizontal line represents $C$.
Although $\lambda_\star=\mathrm{O}(\veps)$, integration over the screening region gives the dimensionless combination
\begin{equation}
    g_\star
    =
    \frac{\lambda_\star h_\star^2}{24\pi^2(\beta m_0)^2}
    =
    \frac{3(N+8)}{\pi^2(N+2)}+\mathrm{O}(\veps).
    \label{eq:gN_Definition}
\end{equation}
Since $g_\star$ remains finite as $\veps\to0$, arbitrarily many quartic vertices contribute at the same order.

To obtain the finite correction to the perimeter coefficient, we first determine the screened field profile sourced by a straight line.
We then evaluate its on-shell action and subtract the Gaussian static contribution already included above.
At leading order in $\veps$, the saddle is rotationally symmetric in the two transverse spatial dimensions.
For $h_\star>0$, we write its nonzero component as
\begin{equation}
    \frac{\langle \phi^1_0(r_\perp)D(C)\rangle_\beta}{\langle D(C)\rangle_\beta}
    =
    \frac{F(z)}{\beta},
    \qquad
    \text{with }
    z=\frac{r_\perp}{\beta},
    \label{eq:StaticLineProfile}
\end{equation}
where $r_\perp$ is the distance from the straight line and $F(z)<0$.
At this order, the profile equation and boundary conditions are
\begin{align}
    F''(z)+\frac{1}{z}F'(z)
    &=(\beta m_0)^2F(z)+\frac{\lambda_\star}{6}F(z)^3,
    \label{eq:RadialProfileEquation}\\
    \lim_{z \to 0} z F'(z) &= \frac{h_\star}{2\pi}\,, \\
    \lim_{z \to \infty} F(z) &= 0.
    \label{eq:RadialProfileBoundaryConditions}
\end{align}
The Gaussian reference at the same $h_\star$ and $m_0$ is $F_0(z)=-h_\star K_0(\beta m_0z)/(2\pi)$, with $K_0$ the modified Bessel function.
Both profiles have the same logarithmic singularity; their finite constants are defined by
\begin{align}
    F(z)&
    \overset{z \to 0}{=}
    \frac{h_\star}{2\pi}\log z+C_\star+ \mathrm{o}(1)\,, \\
    C_0
    &=
    \frac{h_\star}{2\pi}
    \left[
        \log\!\left(\frac{\beta m_0}{2}\right)
        +\gamma_E
    \right]\,,
\end{align}
where $\gamma_{\mathrm E}$ is Euler's constant.
$C_\star$ is a constant that depends however on $N$.
It can be determined for any choice of $N$.
$C_0$ plays the same role for the reference profile.

The Gaussian calculation already includes the static response without quartic interactions.
We therefore compare the nonlinear saddle with a Gaussian saddle at the same source strength and thermal mass, isolating the additional interaction contribution without counting the Gaussian term twice.
Their common short-distance divergence cancels in the difference.
Using a common short-distance regulator $r_0$, the difference of the two on-shell static actions gives the finite correction
\begin{align}
    \frac{\Delta W_{\mathrm{stat}}}{L}
    &=
    \lim_{r_0\to 0}\frac{I_{\mathrm{stat}}^{(0)}[F_0]-I_{\mathrm{stat}}[F]}{L}
    =\frac{h_\star^2}{4\pi\beta}B_N(g_\star),
    \label{eq:NonlinearStaticCorrection}
\end{align}
Here, as explained above, $I_{\mathrm{stat}}$ includes the line source and quartic interaction, and $I_{\mathrm{stat}}^{(0)}$ has the same source and mass but no quartic interaction.
The contribution $B_N(g_\star)$ can be expressed as
\begin{equation}
    B_N(g_\star)
    =-\frac{2\pi}{h_\star} (C_\star-C_0)
    +\frac{\pi^2\lambda_\star}{3h_\star^2}
    \int_0^\infty \mathrm{d}z\,zF(z)^4.
    \label{eq:NonlinearProfileContribution}
\end{equation}
We evaluate $B_N(g_\star)$ by solving the radial equation numerically, extracting $C_\star$, and computing the convergent integral.
The resulting values in Table~\ref{tab:NonlinearProfileContribution} determine the nonlinear contribution at order $\veps^0$.

\begin{table}[t]
    \centering
    \caption{Leading small-$\veps$ values of the finite profile correction $B_N(g_\star)$, the fluctuation coefficient $\alpha_N$, and the perimeter coefficient $m_D$ through order $\sqrt{\veps}$.
    The last row takes $N\to\infty$ after the small-$\veps$ calculation and reports $\alpha_N/N$ and $m_D/N$ in the last two columns.
    The estimated numerical uncertainty of the computed coefficients is $\sim 10^{-5}$, separate from omitted higher-order terms.}
    \begin{tabular}{c c c c}
        \hline
        $N$ & $B_N(g_\star)$ & $\alpha_N$ & $m_D(\veps)$ \\
        \hline
        $1$      & $-0.30371$ & $2.73347$ & $-\frac{9}{4}\log(\veps)-3.13378+2.73347\sqrt{\veps}$ \\
        $2$      & $-0.26715$ & $3.24568$ & $-\frac{5}{2}\log(\veps)-3.75501+3.24568\sqrt{\veps}$ \\
        $3$      & $-0.24349$ & $3.76502$ & $-\frac{11}{4}\log(\veps)-4.35190+3.76502\sqrt{\veps}$ \\
        $4$      & $-0.22688$ & $4.28661$ & $-3\log(\veps)-4.93381+4.28661\sqrt{\veps}$ \\
        $10$     & $-0.18194$ & $7.41354$ & $-\frac{9}{2}\log(\veps)-8.29086+7.41354\sqrt{\veps}$ \\
        $\infty$ & $-0.13093$ & $0.51506$ & $-\frac{1}{4}\log(\veps)-0.53647+0.51506\sqrt{\veps}$ \\
        \hline
    \end{tabular}
    \label{tab:NonlinearProfileContribution}
\end{table}

\paragraph{First fluctuation correction.}
We now compute the order-$\sqrt{\veps}$ correction by integrating fluctuations around the nonlinear static profile.
Keeping terms quadratic in these fluctuations gives a one-loop determinant whose operators depend on the profile.
After renormalization, this static loop contributes an amount of order $m_0$ per unit line length (more precisely: $\beta m_0\sim\sqrt{\veps}$ to the dimensionless coefficient $m_D$).
The determinant includes static loop diagrams with arbitrarily many source insertions, so correcting the bulk propagator in the two-source diagram alone would be insufficient.
Small fluctuations around the line profile are governed by two operators: $\mathcal H_\parallel$ for the sourced field component and $\mathcal H_\perp$ for each of the other $N-1$ components.
They are defined as
\begin{equation}
    \begin{aligned}
        \mathcal H_\parallel
        &= \mathcal H_0 + \frac{\lambda_\star}{2\beta^2} F^2(z), \\
        \mathcal H_\perp
        &= \mathcal H_0 + \frac{\lambda_\star}{6\beta^2} F^2(z),
    \end{aligned}
\end{equation}
with $\mathcal H_0 = -\nabla^2 + m_0^2$, and where $\nabla^2$ acts in three spatial dimensions at this order of the $\veps$-expansion.
Dividing their Gaussian functional integral by the defect-free one gives
\begin{equation}
    \delta m_D
    =
    -\pi\beta \lim_{L\to\infty}\frac{1}{L}
    \left[
        \log \left( \frac{\det\mathcal H_\parallel}{\det\mathcal H_0} \right)
        +
        (N-1) \log \left( \frac{\det\mathcal H_\perp}{\det\mathcal H_0} \right)
    \right].
\end{equation}
The straight line allows us to Fourier transform along it, with momentum $k$.
Because $F(z)$ depends only on the transverse distance to the defect, we may decompose each two-dimensional transverse operator into angular-momentum channels.
For each $k$ and angular momentum, we compute a radial determinant relative to the defect-free operator, sum the channels, and integrate over $k$.
The first-order term in the profile-dependent potential must be treated separately.
With $G_k=(-\nabla_\perp^2+k^2+m_0^2)^{-1}$, the prescription is
\begin{equation}
    \begin{aligned}
    \lim_{L\to\infty}\frac{1}{L}
    \left[\log\left(\frac{\det\mathcal H_i}{\det\mathcal H_0}\right)\right]
    &=
    \int_{-\infty}^{\infty}\frac{\mathrm{d}k}{2\pi}\,
    \operatorname{Tr}_{\perp}
    \left[\log(1+G_k (\mathcal H_i-\mathcal H_0))-G_k (\mathcal H_i-\mathcal H_0)\right] \\
    &\quad
    -\frac{m_0}{4\pi}\int\mathrm{d}^2r_\perp\, (\mathcal H_i-\mathcal H_0),
    \end{aligned}
\end{equation}
where here the index $i$ should be understood as $i=\parallel,\perp$.
The last term is the finite first-order contribution in dimensional regularization.
It already accounts for the soft thermal-mass correction, so we do not add that correction again.
We denote this contribution by
\begin{equation}
    \delta m_D
    =
    \alpha_N \sqrt{\veps}
    +
    \mathrm{o} (\sqrt{\veps})\,.
\end{equation}
Numerical evaluation of the determinants gives the results presented in Table \ref{tab:NonlinearProfileContribution}.

\paragraph{Final result.}
We now add the finite static correction at $g=g_\star$ and the first fluctuation correction to the Gaussian thermal result.
This gives
\begin{equation}
    \lim_{L\to\infty}\frac{W_\beta[C]}{L}
    =
    \frac{h_\star^2}{4\pi\beta}
    \left[
        -\log(\beta m_0)+B_N(g_\star)
    \right]
    +
    \frac{\alpha_N}{2\pi\beta}\sqrt{\veps}
    +
    \frac{o(\sqrt{\veps})}{\beta}.
\end{equation}
Matching to Equation~\eqref{eq:StraightLinePerimeterMatching}, we obtain
\begin{equation}
    m_D
    =
    \frac{N+8}{2}
    \left[
        -\log(\beta m_0)+B_N(g_\star)
    \right]
    +
    \alpha_N\sqrt{\veps}
    +
    o(\sqrt{\veps}).
    \label{eq:WilsonFisherPerimeterCoefficient}
\end{equation}
Using the leading thermal mass, we can write the result as
\begin{equation}\label{eq:epsilonexpresult}
    \begin{aligned}
        m_D
        &=
        -\frac{N+8}{4}\log(\veps)
        -
        \frac{N+8}{4}
        \log\!\left(\frac{2\pi^2(N+2)}{3(N+8)}\right)
        \\
        &\quad
        +
        \frac{N+8}{2}B_N(g_\star)
        +
        \alpha_N\sqrt{\veps}
        +
        o(\sqrt{\veps}).
    \end{aligned}
\end{equation}
Note that the nonlinear profile contributes to the finite term already at leading order.
In Table \ref{tab:NonlinearProfileContribution}, we provide the explicit $\veps$-expanded expression for specific $N$.

\paragraph{$d=2$ and Padé interpolation for $N=1$.}
In $d=2$, the magnetic line cuts the thermal cylinder into a strip of width $\beta$ with identical fixed-spin boundary conditions.
After the temperature-independent perimeter subtraction, comparing the strip and cylinder vacuum energies gives
\begin{equation}
    m_D(d=2)
    =
    2\pi^2 c\left(\frac{1}{24}-\frac{1}{6}\right)
    =
    -\frac{\pi^2}{8},
    \qquad c=\frac{1}{2}.
\end{equation}
To interpolate between $d=4$ and $d=2$, we approximate $m_D+\frac{9}{4}\log\veps$ by a $[1/1]$ Pad\'e in $\sqrt{\veps}$.
Matching the constant and $\sqrt{\veps}$ terms of the expansion above and imposing the exact value at $\veps=2$ gives
\begin{equation}\label{eq:Padèextra}
    m_D^{\mathrm{Pad\acute{e}}}(\veps)
    =
    -\frac{9}{4}\log (\veps)
    +
    \frac{-3.133784+2.473397\sqrt{\veps}}
         {1+0.082991\sqrt{\veps}}.
\end{equation}
At $d=3$, this yields the prediction $m_D^{\mathrm{Pad\acute{e}}} \sim -0.610$.

\subsubsection{The bending coefficient $t_2$}
\label{subsubsec:BendingCoefficient}

To extract $t_2$, we consider a weakly deformed spatial line $C_A$ in flat thermal space at fixed Euclidean time,
\begin{equation}
    \bm{x} (s) = \bigl(s,A\cos(ks),0\bigr),
    \qquad Ak\ll1,
    \qquad k\ll m_0.
    \label{eq:BendingDeformation}
\end{equation}
Here $s$ is distance along the undeformed line, $A$ is the amplitude, and $k$ is the wave number.
We take $A\to0$ and $k/m_0\to0$ before $\veps\to0$, at fixed $N$.
For a projected length $L_0$ containing an integer number of wavelengths, the EFT predicts the quadratic response relative to the straight line $C_{\mathrm{st}}$,
\begin{equation}
    \frac{W_\beta[C_A]-W_\beta[C_{\mathrm{st}}]}{L_0}
    =
    \frac{m_D A^2k^2}{8\pi\beta}
    +
    \frac{\beta t_2 A^2k^4}{4\pi}
    +\cdots.
    \label{eq:BendingMatching}
\end{equation}
The $k^2$ term accounts for the increase in length and is fixed by $m_D$.
Matching the $k^4$ term therefore determines $t_2$.

\paragraph{Gaussian contribution.}
We compute the diagram \eqref{eq:Gaussian_Diagram} with the bent line.
For a fixed separation measured along the line, bending brings the source insertions closer in space and changes the propagator joining them.
Expanding this change to second order in curvature gives the Gaussian bending coefficient:
\begin{equation}
    t_2^{\mathrm{G}}
    =
    \frac{\pi h_\star^2}{8\beta}
    \int_{-\infty}^\infty \mathrm{d}v\,v^2 G(v)
    +o(\veps^0)
    =
    \frac{h_\star^2}{16(\beta m_0)^2}
    +o(\veps^0).
    \label{eq:GaussianBending}
\end{equation}
The nonzero Matsubara modes do not affect the leading $1/\veps$ term.
The factor $m_0^{-2}$ reflects the transverse size of the screening cloud: a broader cloud is more sensitive to bending.
However, as in the perimeter calculation, the finite value of $g_\star$ requires the nonlinear static interactions to be included at leading order.

\paragraph{Nonlinear static contribution.}
We include these interactions by computing the response of the straight-line profile $F(z)$ to a small bend.
This uses the same profile and coordinate $z=r_\perp/\beta$ as in the perimeter calculation.
A uniform transverse displacement simply translates the profile, producing a change proportional to $F'(z)$.
When the displacement varies along the line, the surrounding field must also adjust to the curvature.

We describe this adjustment by a radial function $Y(z)$.
For local curvature $K$ directed along the transverse angle $\vartheta=0$, the first curvature correction to the static field is $-K Y(z)\cos\vartheta$.
Linearizing the field equation about the nonlinear profile gives
\begin{equation}
    \left[
    -\frac{\mathrm{d}^2}{\mathrm{d}z^2}
    -\frac{1}{z}\frac{\mathrm{d}}{\mathrm{d}z}
    +\frac{1}{z^2}
    +(\beta m_0)^2
    +\frac{\lambda_\star}{2}F(z)^2
    \right]Y(z)
    =
    F'(z).
    \label{eq:BendingResponse}
\end{equation}
The $z^{-2}$ term comes from the angular dependence of a transverse displacement, while the term proportional to $F^2$ accounts for the interaction with the screening cloud.
Thus the additional equation is linear even though the background profile is nonlinear.
Keeping the line source fixed and requiring decay far from the defect selects
\begin{equation}
    Y(z)=
    -
    \frac{h_\star}{4\pi} z \log (z)
    +
    \mathrm{O}(z)
    \quad (z\to0),
    \qquad
    Y(z) \to 0
    \quad (z\to\infty).
    \label{eq:BendingResponseBoundary}
\end{equation}
After the contribution from the change in length has been accounted for, the on-shell action expresses the bending coefficient as
\begin{equation}
    t_2
    =
    \pi^2\int_0^\infty \mathrm{d}z\,zF'(z)Y(z)
    +o(\veps^{-1}).
    \label{eq:NonlinearBending}
\end{equation}
The integral measures the overlap between the translation of the original profile and its additional response to curvature.
It is convergent at both endpoints.
Numerically, we therefore reuse the straight-line solution, solve one linear radial equation, and evaluate this integral.
This expression includes the Gaussian contribution: inserting $F_0(z)=-h_\star K_0(\beta m_0z)/(2\pi)$ gives $Y_0(z)=h_\star zK_0(\beta m_0z)/(4\pi)$ and reproduces~\eqref{eq:GaussianBending} at leading order.

\paragraph{Final result.}
The leading bending coefficient takes the form
\begin{equation}
    t_2=\frac{c_N}{\veps}+o(\veps^{-1}),
    \label{eq:BendingFinal}
\end{equation}
with the values of $c_N$ listed in Table~\ref{tab:BendingCoefficient}.
The nonlinear response reduces the Gaussian estimate substantially, confirming that the same static interactions needed for the perimeter coefficient are also essential for bending.

\begin{table}[t]
    \centering
    \begin{tabular}{c c c}
        \hline
        $N$ & $c_N^{\mathrm{G}}$ & $c_N$ \\
        \hline
        $1$ & $0.256469$ & $0.138484$ \\
        $2$ & $0.237472$ & $0.137326$ \\
        $3$ & $0.229872$ & $0.139044$ \\
        \hline
    \end{tabular}
    \caption{Leading coefficients $c_N=\lim_{\veps\to0}\veps t_2$ in the manuscript convention.
    The Gaussian comparison keeps the same source strength and thermal mass.}
    \label{tab:BendingCoefficient}
\end{table}

In our convention, $t_2>0$ means that bending increases $W_\beta$ after the perimeter contribution has been accounted for.
The result determines the leading small-$\veps$ coefficient; fluctuations about the saddle are expected to give relative corrections of order $\sqrt{\veps}$ and are not included here.
The expansion remains valid when the deformation wavelength is much larger than the screening length.

\subsection{$3d$ Ising model}
The $3d$ Ising model cannot be treated analytically, if not as a $\varepsilon\to 1$ limit of the $\varepsilon-$expansion discussed above or a $N\to 1$ limit of a proper large $N$ expansion. None of these options is supposed to be a good approximation and for this reason we are forced to study the model numerically and simulate it on a lattice.\footnote{We will show at the end of the section that even though the $\varepsilon\to 1$ limit is uncontrolled it still gives good estimations.} 

\subsubsection{Lattice model and defect observable}
\label{subsubsec:LatticeModelAndDefectObservable}

We study the critical three-dimensional Ising model on a periodic isotropic
cubic lattice,
\begin{equation}
  S^1_\beta\times T^2,
  \qquad
  N_\tau=\frac{\beta}{a},
\end{equation}
with spins $s_x=\pm1$ and bulk partition function
\begin{equation}
  Z_0=
  \sum_{\{s\}}
  \exp\left[
    K_c\sum_{\langle xy\rangle}s_xs_y
  \right],
  \qquad
  K_c=0.221654626.
  \label{eq:bulk-critical-partition-function}
\end{equation}
The coupling is the flat-space critical Ising coupling (computed e.g. in \cite{Ferrenberg_2018,PhysRevE.101.022131}) and is held fixed
throughout the calculation.  Compactifying Euclidean time does not entail a
new tuning of the bulk critical point. Our purpose is to check the EFT in equation \eqref{eq:LogPartitionFunction} and predict the first coefficients in the high temperature expansion. For this purpose the fact that the spatial manifold is not a sphere as in the derivation of the asymptotic is not important: the EFT has to be valid in any manifold in the high temperature limit.

The defect is a closed magnetic line $C$, placed at one Euclidean-time slice
and rasterized as a spatial circle of radius $R$.  Its physical definition is
the fixed-spin operator
\begin{equation}
  {\cal D}_{\rm pin}(C)
  =
  \prod_{i\in C}\delta_{s_i,+1}.
  \label{eq:pinned-defect}
\end{equation}
It constrains every spin on the loop to be up.  This is the strong-field
representative of the symmetry-breaking magnetic line fixed point.  It is a
local codimension-two defect: no sheet of modified bonds, frustrated plaquette,
or change of bulk holonomy is introduced.

For the free-energy measurement we introduce an auxiliary line field,
\begin{equation}
  {\cal D}_h(C)
  =
  \exp\left[
    h\sum_{i\in C}(s_i-1)
  \right],
  \qquad
  {\cal D}_{\rm pin}(C)
  =
  \lim_{h\to+\infty}{\cal D}_h(C).
  \label{eq:auxiliary-pinning-field}
\end{equation}
The subtraction by $1$ is essential: a spin already fixed to $+1$ contributes
unit weight, while a down spin costs a factor $e^{-2h}$.  Thus $h$ is only an
auxiliary thermodynamic-integration coordinate. 

Define
\begin{equation}
  Z_h(C)
  =
  \sum_{\{s\}}
  \exp\left[
    K_c\sum_{\langle xy\rangle}s_xs_y
    +h\sum_{i\in C}(s_i-1)
  \right],
  \qquad
  W_h(C)=\log\frac{Z_h(C)}{Z_0}.
  \label{eq:pinning-partition-function}
\end{equation}
Since $Z_{h=0}=Z_0$, differentiation gives
\begin{align}
  \frac{\partial W_h}{\partial h}
  =
  \left\langle\sum_{i\in C}(s_i-1)\right\rangle_h \nonumber=
  \langle M_C\rangle_h-|C|
  =
  -2\langle N_\downarrow(C)\rangle_h,
  \label{eq:pinning-integrand}
\end{align}
where $M_C=\sum_{i\in C}s_i$ and $N_\downarrow(C)$ counts down spins on the
loop.  The desired defect one-point function is therefore obtained directly
from
\begin{equation}
  W_{\rm pin}(C)
  \equiv
  \log\frac{Z_{\rm pin}(C)}{Z_0}
  =
  \int_0^\infty dh\,
  \left[
    \langle M_C\rangle_h-|C|
  \right].
  \label{eq:pinning-thermodynamic-integration}
\end{equation}
This formula already divides by the bulk partition function.
Hence it measures the logarithm of the normalized defect one-point function,
not the global bulk free energy.

The integral is evaluated up to a sufficiently large $h_{\max}$.  The omitted
tail is exponentially bounded. In fact it is easy to show that 
\begin{equation}
  \left|W_{\rm pin}-W_{h_{\max}}\right|
  \leq
  |C|e^{-2(h_{\max}-6K_c)}.
  \label{eq:pinning-tail-bound}
\end{equation}
The final uncertainty combines Monte-Carlo statistics,  and this tail estimate.

\subsubsection{Perimeter coefficient for the straight line}
\label{subsubsec:ScalingTestForTheCircularLine}

\begin{figure*}[t!]
    \centering
    \begin{subfigure}[b]{0.5\textwidth}
        \centering
        \includegraphics[width=0.95\textwidth]{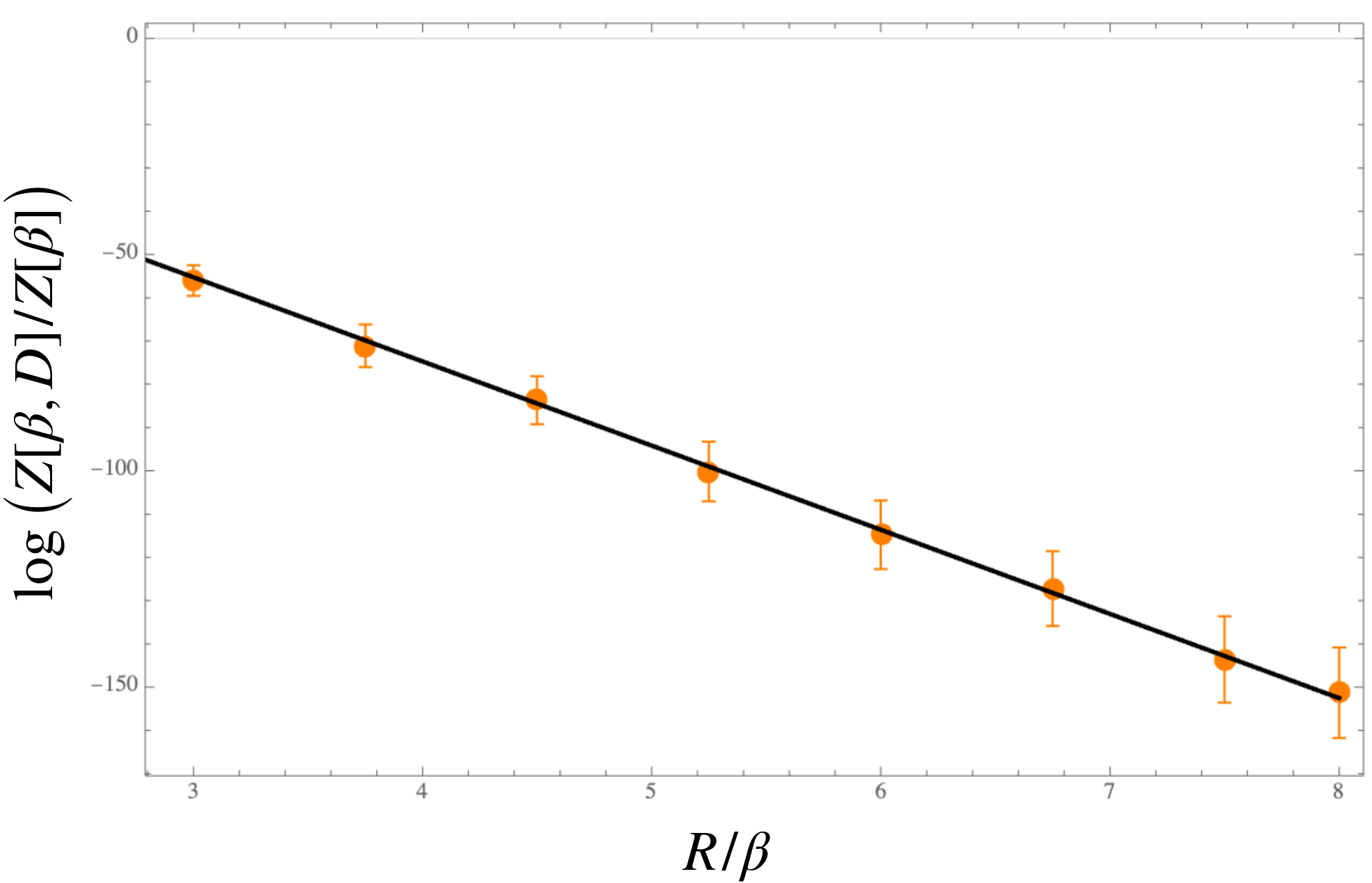}
    \end{subfigure}%
    ~ 
    \begin{subfigure}[b]{0.5\textwidth}
        \centering
        \includegraphics[width=0.95\textwidth]{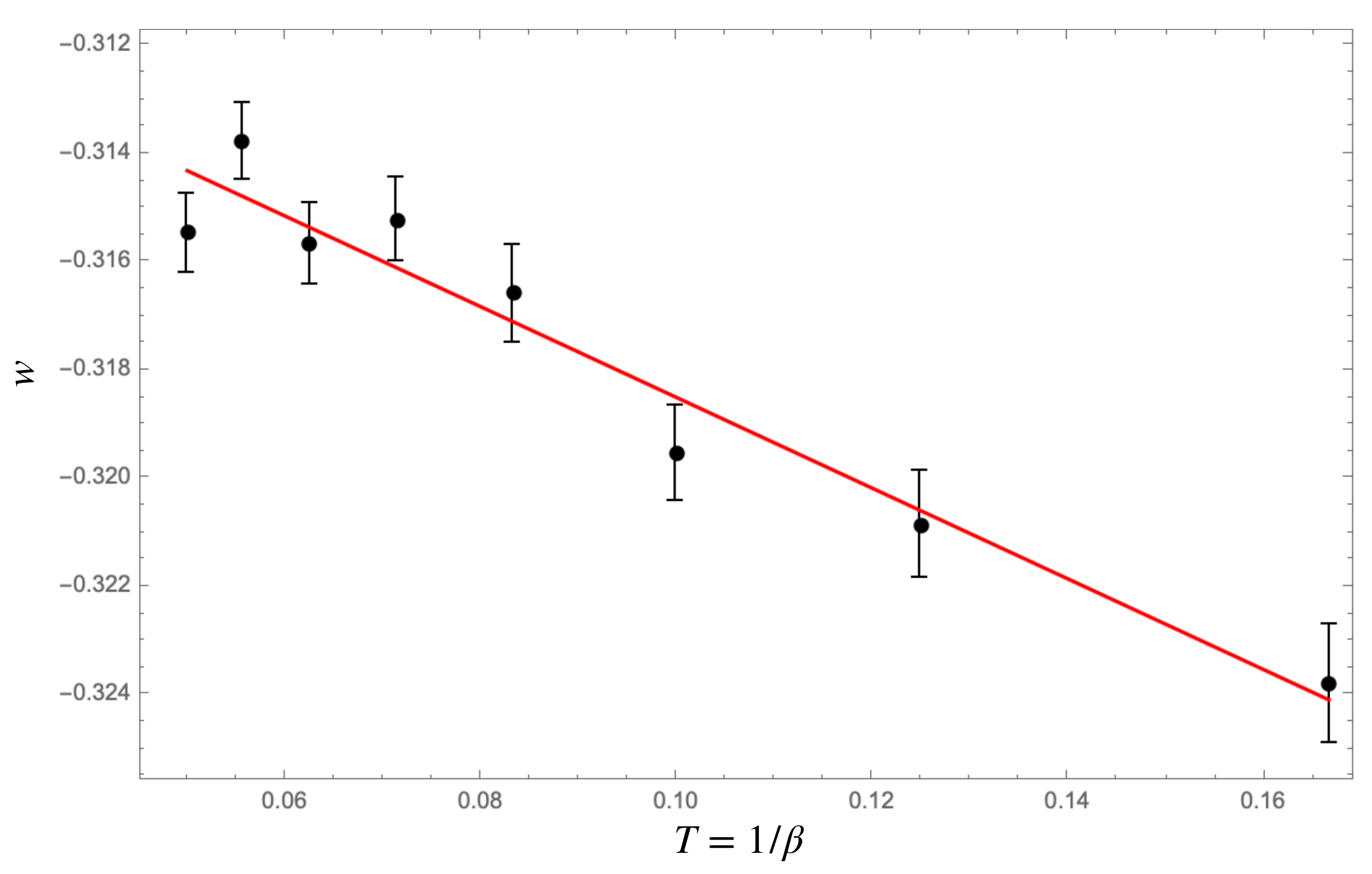}
    \end{subfigure}
    \caption{\textbf{Left Panel}: The partition-function ratio
    $\log\!\bigl(Z[\beta,\mathcal D]/Z[\beta]\bigr)$
    in the three-dimensional Ising model as a function of $R/\beta$.
    Orange points show the Monte Carlo data, and the black line shows
    the fit motivated by the thermal EFT. \textbf{Right panel}: Fit of $W_{\rm pin}/N_x$ in $1/\beta$: the slope of the linear fit is expected to be $2 \pi m_D$.}
    \label{fig:double}
\end{figure*}

Further details of the simulations are given in Appendix~\ref{app:details}. The results are shown in the left panel of  Figure~\ref{fig:double}.

At large $R/\beta$, the data are consistent with the leading EFT
prediction. Writing $\rho=R/\beta$ for the geometry considered here,
this prediction takes the form
\begin{equation}
    \log\frac{Z[\beta,\mathcal D]}{Z[\beta]}
    =m_D^{\rm bare}\rho+\gamma+\ldots\,.
\end{equation}
And we should substract from the coefficient $m_D^{\rm bare}$ the scheme dependent contribution  contribution, that could be obtained by studying ths system at zero temperature. 
A fit to the simulation data gives
\begin{equation}
    m_D^{\rm bare}=-19\pm2\,,
\end{equation}
with a reduced chi-squared of
$\chi^2/\mathrm{dof}\simeq0.025$.\footnote{We also tested a fit
including an area term proportional to $(R/\beta)^2$. Its coefficient
is consistent with zero, and the reduced chi-squared increases to
$\chi^2/\mathrm{dof}\simeq0.177$.}
Thus we need to find the temperature dependent part.
We therefore performed a new, eight-point Monte Carlo scan for various \(N_t= \beta\) \footnote{Specifically $N_t = 6,8,10,12,14,16,18,20$.}, placing a straight, spatially wrapping pinned line along the \(x\) direction in boxes with \(N_x=8 \beta\) and \(N_y=12\beta\). At each temperature we simulated various pinning fields between \(h=0\) and \(h=8\). We used Bennett's acceptance-ratio method \footnote{The Bennett acceptance-ratio method estimates the free-energy difference between two nearby field values by combining Monte Carlo samples from both ensembles.
We applied the latter method successively between adjacent fields to obtain the total free-energy change, avoiding numerical integration of the magnetization.} to determine the free-energy difference between neighboring fields from the measured line magnetization, thereby avoiding a numerical quadrature of the field-dependent expectation value. After subtracting the known large-field normalization, we obtained the free energy per unit line length, \(w(\beta)=W_{\rm pin}(\beta)/N_x\), and fitted it to 
\[w(N_t)=-\kappa+\frac{m_D}{2\pi \beta}+O(\beta^{-2}) \ .\] 
The weighted linear fit yields 
The plot compared with the fit is shown in the right panel of Figure~\ref{fig:double}.
\[m_D=-0.53\pm0.06 \ .\]
 Longer chains and further finite-size studies are still needed for a controlled continuum estimate more accurately but we leave them for future studies.

\begin{figure*}[t]
    \centering
    \includegraphics[width=0.8\textwidth]{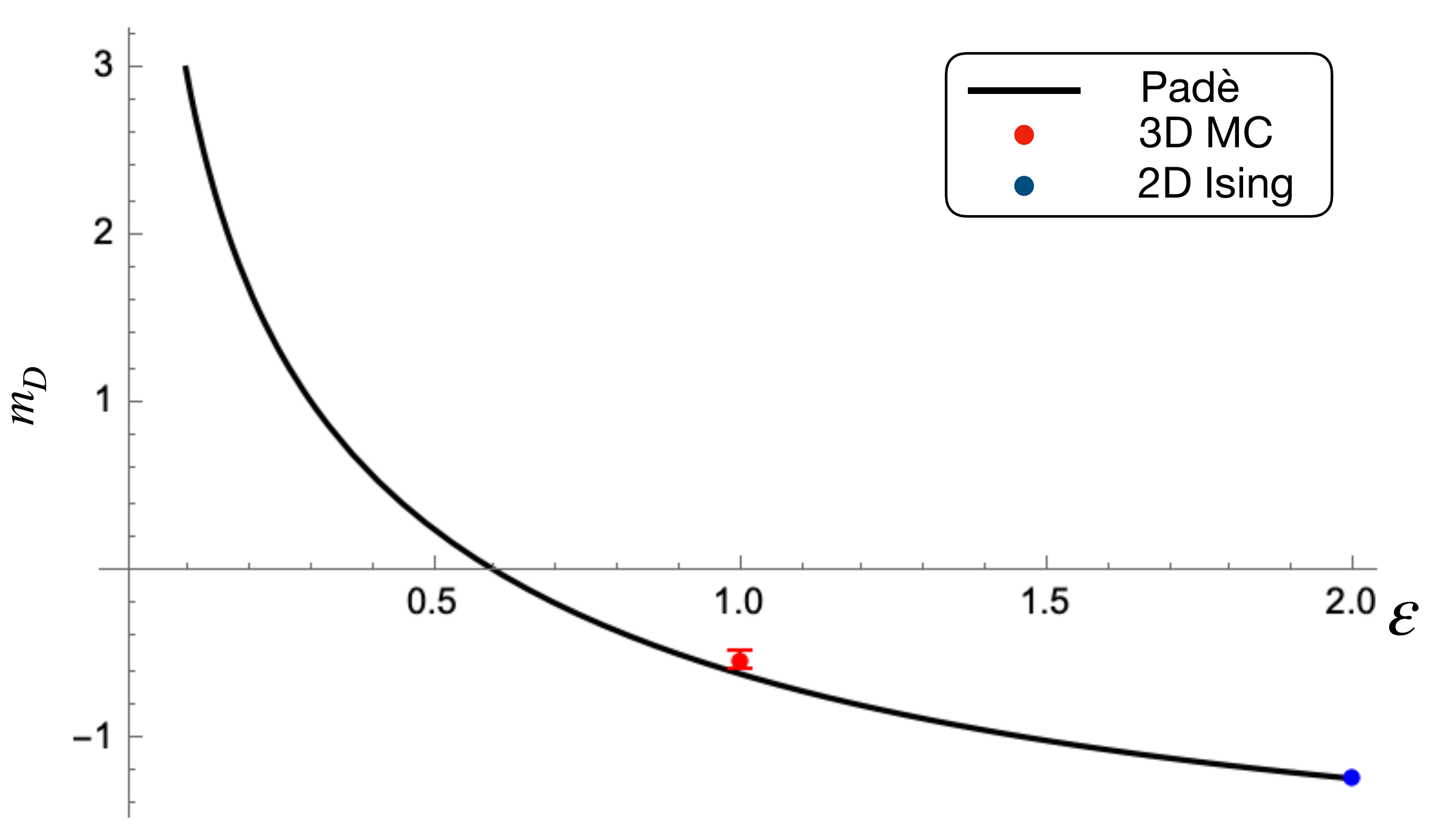}
    \caption{The Padé extrapolation for the value of $m_D$ as a function of $\varepsilon$ compared with Monte Carlo result for $3d$ Ising (red point) and exact result in 2D Ising (blue dot).}
    \label{Fig:Pade}
\end{figure*}

Finally, let us compare the Monte Carlo prediction with the Padé extrapolation from the $\varepsilon$-expansion. The function of $\varepsilon$ is given in equation \eqref{eq:Padèextra} and plotted in Figure \ref{Fig:Pade}. In $3d$ the approximations gives $m_D \sim -0.61$ which, considering solvely the statistical error from Monte Carlo, it matches up to  $\sim 1.6 \sigma$, conferming the prediction.

\subsection{Large-$N$ Gross--Neveu model}
\label{subsec:GrossNeveu}
\subsubsection{Setup}
\label{subsubsec:GNSetup}

\paragraph{Action.}
We consider $N$ two-component Dirac fermions in $d=3$ with the action
\begin{equation}
    S
    =
    \int\mathrm{d}^3x
    \left[
        \bar\psi_i\left(\slashed\partial+\sigma\right)\psi_i
        +\frac{N}{2g}\,\sigma^2
    \right],
    \qquad
    i=1,\ldots,N,
\end{equation}
at the critical coupling.
Since $\det(i\slashed Q+\sigma)=Q^2+\sigma^2$, integrating out the fermions gives
\begin{equation}
    \frac{1}{g_c}
    =
    2\int\!\frac{\mathrm{d}^3Q}{(2\pi)^3}\,\frac{1}{Q^2}
    =
    \int\!\frac{\mathrm{d}^2q}{(2\pi)^2}\,\frac{1}{q}.
\end{equation}
The field $\sigma$ is odd under the discrete symmetry that forbids a fermion mass.
At large $N$, its dimension is $\Delta_\sigma=1+\mathrm{O}(1/N)$, with a negative $1/N$ correction.

\paragraph{Line.}
We define $\hat\sigma=\sqrt N\,\sigma$ and the unit-normalized operator $O=\frac{\pi}{2}\hat\sigma$, and consider
\begin{equation}
    D(C)
    =
    \exp\!\left[
        -h\int_C\mathrm{d}\ell\,O
    \right].
\end{equation}
The connected $n$-point functions of $\hat\sigma$ scale as $N^{1-n/2}$.
At $N=\infty$, $O$ is therefore a Gaussian field of dimension one, and $D(C)$ is a conformal line for every $h$.

\subsubsection{Thermal propagator}
\label{subsubsec:GNPropagator}

\paragraph{Fermion bubble.}
At $N=\infty$, the propagator of $\hat\sigma$ is the inverse of the fermion bubble,
\begin{equation}
    \langle\hat\sigma\hat\sigma\rangle_\beta(P)
    =
    \frac{1}{\Pi_\beta(P)},
    \qquad
    \Pi_\beta(P)
    =
    \frac{1}{g_c}
    -
    \begin{tikzpicture}[scale=0.9, baseline={([yshift=-.8ex]current bounding box.center)}]
        \draw[scalar] (-.9,0) -- (-.45,0);
        \draw[scalar] (.45,0) -- (.9,0);
        \draw[thick,-{Latex[length=1.6mm]}] (-.45,0) arc (180:90:.45);
        \draw[thick] (0,.45) arc (90:0:.45);
        \draw[thick,-{Latex[length=1.6mm]}] (.45,0) arc (360:270:.45);
        \draw[thick] (0,-.45) arc (270:180:.45);
        \draw[VertexColor,fill=VertexColor] (-.45,0) circle (.065);
        \draw[VertexColor,fill=VertexColor] (.45,0) circle (.065);
    \end{tikzpicture}\,.
\end{equation}
Using $\mathrm{tr}[\slashed Q(\slashed Q+\slashed P)]=2Q\cdot(Q+P)$ and $2Q\cdot(Q+P)=Q^2+(Q+P)^2-P^2$, the bubble becomes
\begin{equation}
    \Pi_\beta(P)
    =
    \frac{1}{g_c}
    -\int_Q\left[\frac{1}{Q^2}+\frac{1}{(Q+P)^2}\right]
    +P^2J_\beta(P),
    \qquad
    J_\beta(P)
    =
    \int_Q\frac{1}{Q^2(Q+P)^2}.
\end{equation}
Here $\int_Q\equiv\frac1\beta\sum_m\int\frac{\mathrm{d}^2q}{(2\pi)^2}$, with $Q=(\nu_m,\bm q)$ and fermionic frequencies $\nu_m=(2m+1)\pi/\beta$, while $P=(\omega_n,\bm k)$ with $\omega_n=2\pi n/\beta$.
A bosonic shift preserves the fermionic frequencies, so both tadpoles equal to  \[\int\frac{\mathrm{d}^2q}{(2\pi)^2}\frac{\tanh(\beta q/2)}{2q} \ .\]
Subtracting $1/g_c$, we obtain
\begin{equation}
    \Pi_\beta(P)
    =
    \frac{\log 2}{\pi\beta}+P^2J_\beta(P).
    \label{eq:GNBubble}
\end{equation}
At zero temperature, $J=1/(8|P|)$ and the propagator reduces to $\langle\hat\sigma\hat\sigma\rangle(P)=8/|P|$, i.e.\ $4/(\pi^2x^2)$ in position space.

\paragraph{Screening.}
At $g_c$, the finite-temperature gap equation $1/g_c=2\int\tanh(\beta E/2)/(2E)$, with $E=\sqrt{q^2+\bar\sigma^2}$, admits only $\bar\sigma=0$.
The thermal state is therefore symmetric.
Expanding~\eqref{eq:GNBubble} at small momentum gives
\begin{equation}
    \Pi_\beta(0,\bm k)
    =
    \frac{\log2}{\pi\beta}
    +\frac{\beta\,\bm k^2}{16\pi}
    +\ldots,
\end{equation}
which defines a static screening scale $\sqrt{16\log2}/\beta\approx3.3/\beta$.
Unlike at the Wilson--Fisher fixed point, there is no parametrically light static mode.

\paragraph{The integral $J_\beta$.}
Combining the denominators with a Feynman parameter $x$, integrating over $\bm q$, and using the fermionic sum
\begin{equation}
    \frac1\beta\sum_m\frac{1}{(\nu_m+a)^2+b^2}
    =
    \frac{\sinh(\beta b)}{2b\left[\cosh(\beta b)+\cos(\beta a)\right]},
\end{equation}
we find
\begin{equation}
    J_\beta(\omega_n,\bm k)
    =
    \frac{1}{4\pi}\int_0^1\mathrm{d}x\,
    \frac{\sinh(\beta b)}{2b\left[\cosh(\beta b)+\cos\big(2\pi n(1-x)\big)\right]},
    \qquad
    b=\sqrt{x(1-x)}\,|P|.
    \label{eq:GNJ}
\end{equation}
At zero temperature this reproduces $J=1/(8|P|)$.
We have also checked~\eqref{eq:GNJ} against a direct Matsubara sum.

\paragraph{Endpoint layer.}
Substituting $x=\sin^2\theta$ and using the symmetry $\theta\to\pi/2-\theta$, we can write~\eqref{eq:GNBubble} as
\begin{equation}
    \Pi_\beta(P)
    =
    \frac{|P|}{8}
    +\frac{1}{\beta}\left[
        \frac{\log2}{\pi}
        +\frac{\beta|P|}{2\pi}\int_0^{\pi/4}\mathrm{d}\theta\,
        \big(F(\theta)-1\big)
    \right],
    \qquad
    F=\frac{\sinh(\beta b)}{\cosh(\beta b)+\cos\big(2\pi n\cos^2\theta\big)}.
    \label{eq:GNEndpoint}
\end{equation}
The integrand is exponentially small unless $\theta\lesssim1/(\beta|P|)$.
In this layer $F-1\to\tanh(\varphi/2)-1$, with $\varphi=\beta|P|\theta$, and $\int_0^\infty\mathrm{d}\varphi\,[\tanh(\varphi/2)-1]=-2\log2$.
Hence the constant term cancels at large momentum, and $\Pi_\beta(P)-|P|/8=\mathrm{O}(\beta^{-2}|P|^{-1})$.

\subsubsection{Leading term in the EFT: $m_D$}
\label{subsubsec:GNPerimeter}

\paragraph{Line integral.}
For a line along $y$, we need to make use of 
\begin{equation}
    \int\mathrm{d}v\,\langle\hat\sigma(v)\hat\sigma(0)\rangle_\beta
    =
    \frac{1}{\beta}\sum_n\int\frac{\mathrm{d}k}{2\pi}\,
    \frac{1}{\Pi_\beta(\omega_n,k)}.
\end{equation}
To organize the zero-temperature subtraction, we add and subtract the reference propagator $8/\sqrt{P^2+M^2}$ for an arbitrary $M>0$.
Its frequency sum can be performed exactly by Poisson resummation, using $G_{\mathrm{ref}}(r)=4MK_1(Mr)/(\pi^2r)$ and $\int\mathrm{d}v\,K_1(M\sqrt{v^2+a^2})/\sqrt{v^2+a^2}=\pi e^{-Ma}/(Ma)$.
We obtain
\begin{equation}
    I_{\hat\sigma}
    =
    \frac1\beta\sum_n\int\frac{\mathrm{d}k}{2\pi}
    \left[
        \frac{1}{\Pi_\beta(\omega_n,k)}
        -\frac{8}{\sqrt{\omega_n^2+k^2+M^2}}
    \right]
    -\frac{8}{\pi\beta}\log\left(1-e^{-\beta M}\right)
    -\frac{4M}{\pi}.
    \label{eq:GNLineIntegral}
\end{equation}
The bracket decays as $n^{-2}$, with the coefficient fixed by the large-momentum expansion.
Independence of the result from $M$ provides a check of the numerics.

\paragraph{Final result.}
We evaluate~\eqref{eq:GNLineIntegral} numerically, using~\eqref{eq:GNEndpoint} for $\Pi_\beta$.
The results are listed in Table~\ref{tab:GNLineIntegral}.
For the unit-normalized operator, $I_O=\frac{\pi^2}{4}I_{\hat\sigma}$. We get therefore
\begin{equation}
    m_D
    =
    \pi\beta\,h_\star^2\,I_O
    =
    -8.3512\,h_\star^2
    +\mathrm{O}(1/N).
    \label{eq:GNResult}
\end{equation}
The perimeter coefficient is negative.
Since the fermions are antiperiodic, the $\sigma$ correlator is screened on scales of order $\beta$, and thermal decorrelation along the line dominates.
This is the same mechanism that makes $m_D$ negative for every weakly coupled line in $d=2$.
It is opposite to the Wilson--Fisher magnetic line near $d=4$, where a static mode of mass $m_0\sim\sqrt\veps/\beta$ makes $I_J$ logarithmically large and positive.

\begin{table}[t]
    \centering
    \caption{Line integral~\eqref{eq:GNLineIntegral} for three values of the auxiliary mass $M$, and the resulting perimeter coefficient~\eqref{eq:GNResult}.
    The estimated numerical uncertainty is $\sim10^{-6}$.}
    \begin{tabular}{c c c c}
        \hline
        $\beta M$ & $\beta I_{\hat\sigma}$ & $\beta I_O$ & $m_D/h_\star^2$ \\
        \hline
        $2$ & $-1.077361$ & $-2.658283$ & $-8.3512$ \\
        $3$ & $-1.077361$ & $-2.658283$ & $-8.3512$ \\
        $5$ & $-1.077362$ & $-2.658283$ & $-8.3512$ \\
        \hline
    \end{tabular}
    \label{tab:GNLineIntegral}
\end{table}
\subsection{$\mathcal{N}=2$ Chern--Simons-matter theory}
\label{subsec:N2ChernSimons}

In every integer-dimensional example above, the perimeter coefficient is negative.
This is not a consequence of unitarity and screening: in $d=3$, the $\mathcal{N}=2$ supersymmetric $U(1)_k$ Chern--Simons theory with one charged chiral multiplet has a conformal line with $m_D>0$.
At large $k$ the line is described by a classical field, and the sign of $m_D$ follows from an exact identity.

\subsubsection{Setup}
\label{subsubsec:N2Setup}

\paragraph{Action.}
We consider a complex scalar $Q$ and a two-component Dirac fermion $\psi$ of unit charge, coupled to a $U(1)$ gauge field with Chern--Simons level $k\in\mathbb{Z}+\tfrac12$.
In Lorentzian signature, and in the conventions of~\cite{Avdeev:1992jt},
\begin{equation}
    \mathcal{L}
    =
    \frac{k}{4\pi}\epsilon^{\mu\nu\rho}A_\mu\partial_\nu A_\rho
    +|D_\mu Q|^2
    +i\bar\psi\gamma^\mu D_\mu\psi
    +y\,\bar QQ\,\bar\psi\psi
    -g\,(\bar QQ)^3,
    \qquad
    D_\mu=\partial_\mu-iA_\mu.
    \label{eq:N2Action}
\end{equation}
The scaling dimensions are $\Delta_Q=\tfrac12$ and $\Delta_\psi=\Delta_A=1$, so that $\kappa\equiv2\pi/k$, $y$ and $g$ are dimensionless.
The theory is $\mathcal{N}=2$ supersymmetric for~\cite{Gaiotto:2007qi,Avdeev:1992jt}
\begin{equation}
    y=3\kappa,
    \qquad
    g=\kappa^2.
    \label{eq:N2SusyPoint}
\end{equation}
These values arise from the auxiliary fields $\sigma$, $\lambda$, $\mathsf{D}$ of the vector multiplet, of dimensions $1$, $\tfrac32$, $2$: the field $\mathsf{D}$ multiplies $\frac{k}{2\pi}\sigma+\bar QQ$ and imposes $\sigma=-\kappa\,\bar QQ$, so that the potential $\sigma^2\,\bar QQ$ becomes $\kappa^2(\bar QQ)^3$, while $\sigma\bar\psi\psi$ and the elimination of $\lambda$ contribute $\kappa$ and $2\kappa$ to $y$.

\paragraph{Conformality.}
At two loops, for generic $y$ and $g$, the beta-functions have already been computed to be equal to
\begin{equation}
    \begin{aligned}
        (4\pi)^2\,\mu\frac{\mathrm{d}y}{\mathrm{d}\mu}
        &=
        \tfrac{14}{3}\,y^3-34\,y\kappa^2-24\,\kappa^3,
        \\
        (4\pi)^2\,\mu\frac{\mathrm{d}g}{\mathrm{d}\mu}
        &=
        168\,g^2-84\,g\kappa^2+16\,gy^2-4\,y^4+4\,y^2\kappa^2+8\,y\kappa^3+36\,\kappa^4,
    \end{aligned}
    \label{eq:N2BetaFunctions}
\end{equation}
One can check that the supersymmetric point is the fixed point of these equations.
The theory is thus a unitary CFT for every $k$.
At large $k$ it is weakly coupled: for $A_\mu=0$ and real $F = \sqrt{\frac{2\sqrt{3}\pi}{k}} Q$, the Euclidean action of the scalar is
\begin{equation}
    S
    =
    \frac{k}{\sqrt{3}\pi}\int\mathrm{d}^3x
    \left[
        \tfrac12(\partial F)^2+\tfrac16F^6
    \right]
    \label{eq:N2ClassicalAction}
\end{equation}
so that $1/k$ is the loop-counting parameter.

\paragraph{Line.}
We consider the following deformation
\begin{equation}
    D_h(C)
     = \exp\!\left[
        h\int_C\mathrm{d}\ell\,\bar QQ
    \right]
    =
    \exp\!\left[
        \frac{hk}{2\sqrt{3}\pi}\int_C\mathrm{d}\ell\,F^2
    \right],
    \qquad
    h>0.
    \label{eq:N2Line}
\end{equation}

Since $\Delta_{\bar QQ}=1$, the coupling $h$ is dimensionless.\footnote{This dimension is exact: $\bar QQ=-\sigma/\kappa$, and $\sigma$ is the scalar in the multiplet of the conserved current $\epsilon^{\mu\nu\rho}\partial_\nu A_\rho$.}
Thus in the presence of the defect we must minize the following quantity
\begin{equation}
       \qquad
    \mathcal{E}[F]
    =
    \int\mathrm{d}^3x
    \left[
        \tfrac12(\partial F)^2+\tfrac16F^6
    \right]
    -\frac{h}{2}\int_C\mathrm{d}\ell\,F^2,
    \label{eq:N2Saddle}
\end{equation}
\paragraph{Fixed point of the line.}
At zero temperature the saddle depends only on the distance $r$ from the line.
Regularizing the source on a tube of radius $a$, it obeys
\begin{equation}
    (rF')'=rF^5
    \quad(r>a),
    \qquad
    \frac{aF'(a)}{F(a)}=-\frac{h}{2\pi},
    \qquad
    F(\infty)=0.
    \label{eq:N2RadialEquation}
\end{equation}
Let $h(r)\equiv-2\pi rF'(r)/F(r)$ be the coupling seen at distance $r$, with $h(a)=h$, and $\varrho(r)\equiv\sqrt{r}\,F(r)$.
The field equation is equivalent to the beta functions
\begin{equation}
    \frac{\mathrm{d}h}{\mathrm{d}\log r}
    =
    \frac{h^2}{2\pi}-2\pi\varrho^4,
    \qquad
    \frac{\mathrm{d}\varrho}{\mathrm{d}\log r}
    =
    \varrho\left(\frac12-\frac{h}{2\pi}\right).
    \label{eq:N2LineBeta}
\end{equation}
The term $h^2/2\pi$ is the running of a $\delta$-function source in two transverse dimensions.
Alone, it would produce a Landau pole at $r=a\,e^{2\pi/h}$, reflecting the fact that $\mathcal{E}$ is unbounded below for a free field.
The sextic interaction stops the growth, and both beta functions vanish at
\begin{equation}
    h_\star=\pi,
    \quad
    \varrho_\star^4=\frac14
    \qquad\Longleftrightarrow\qquad
    F_\star(r)=\frac{1}{\sqrt{2r}},
    \quad
    \langle\bar QQ\rangle=\frac{k}{4\sqrt3\,\pi\,r}.
    \label{eq:N2LineFixedPoint}
\end{equation}
Linearizing~\eqref{eq:N2LineBeta}, deviations scale as $r^s$ with $s^2-s-1=0$.
The growing mode, $s=\tfrac12(1+\sqrt5)$, is excluded for a solution that is regular for all $r$, and the other one decays as $r^{-(\sqrt5-1)/2}$.\footnote{Eliminating $h$ gives $\ddot\varrho-\dot\varrho+\tfrac14\varrho=\varrho^5$, with dots denoting $\mathrm{d}/\mathrm{d}\log r$, for which $\tfrac12\dot\varrho^2+\tfrac18\varrho^2-\tfrac16\varrho^6$ is non-decreasing.
A positive solution that exists for all $r>a$ is bounded, since $\varrho>\varrho_\star$ with $\dot\varrho\geq0$ leads to a singularity at finite $r$.
It therefore converges to a fixed point, which cannot be the repelling node $\varrho=0$.
Moreover, $\dot h=2\pi q\,(\frac{h}{2\pi}+\varrho^2)$ with $q=\frac{h}{2\pi}-\varrho^2$, and $\dot q=q\,(\frac{h}{2\pi}+\varrho^2)+\varrho^2(\frac{h}{\pi}-1)$.
At a last zero of $q$, $\dot q$ would have the sign of $h-\pi$, opposite to the sign of $q$ afterwards.
Hence $q$ does not vanish and $h(r)$ is monotonic, ranging over $(0,\pi)$ on one solution and over $(\pi,\infty)$ on the other.}
Hence $h(r)\to\pi$ monotonically for every $h>0$, and as $a\to0$ all the lines $D_h$ become a single line $D_\star$, defined by $F-F_\star\to0$ as $r\to0$.
The profile $F_\star$ is invariant under the inversion $x^\mu\to x^\mu/x^2$ about a point of the line, under which $F(x)\to|x|^{-1}F(x/x^2)$ and $r\to r/x^2$, and therefore under the full conformal group of the line.
Fluctuations obey $(-\nabla^2+\frac{5}{4r^2})\,\delta F=0$ and behave as $r^{\pm\nu_m}$ with $\nu_m=(m^2+\tfrac54)^{1/2}$ for transverse spin $m$, so the corresponding defect operators have dimensions $\hat\Delta_m=\tfrac12+\nu_m$.
The lowest one, $\hat\Delta_0=\tfrac12(1+\sqrt5)$, is irrelevant, in agreement with the slope $1-\hat\Delta_0$ of the beta function at $h_\star$, and $\hat\Delta_1=2$ is the displacement operator.
These statements hold at leading order in $1/k$.

\subsubsection{Leading EFT coefficient: $m_D$}
\label{subsubsec:N2Perimeter}

\paragraph{Energy formula.}
The line at $\tau=0$ acts as an operator $\hat D$ on the Hilbert space of the spatial slice, and $Z[\beta,D]=\operatorname{Tr}(\hat D\,e^{-\beta H})$.
Neither $\hat D$ nor its renormalization depends on $\beta$.
Differentiating $W_\beta[C]$ with respect to $\beta$ and using~\eqref{eq:StraightLinePerimeterMatching}, we obtain the exact relation
\begin{equation}
    m_D
    =
    \frac{2\pi\beta^2}{L}\,\Delta E,
    \qquad
    \Delta E
    =
    \frac{\operatorname{Tr}(\hat D\,H\,e^{-\beta H})}{\operatorname{Tr}(\hat D\,e^{-\beta H})}
    -
    \frac{\operatorname{Tr}(H\,e^{-\beta H})}{\operatorname{Tr}(e^{-\beta H})}.
    \label{eq:N2EnergyFormula}
\end{equation}
The sign of $m_D$ is the sign of the energy that the line adds to the thermal state.
Since $H$ commutes with $e^{-\beta H}$, this energy can be measured on any slice $\tau=\tau_0$ with $0<\tau_0<\beta$.

\paragraph{Saddle point and sign.}
At leading order in $1/k$, $\Delta E$ is the Euclidean energy of the saddle $F(y,\tau)$ of~\eqref{eq:N2Saddle} for a straight line on $S^1_\beta\times\mathbb{R}^2$, where $y$ is the spatial coordinate transverse to the line:
\begin{equation}
    \Delta E
    =
    \frac{k L}{\sqrt{3\pi} }\int_{-\infty}^{\infty}\mathrm{d}y
    \left[
        -\tfrac12(\partial_\tau F)^2+\tfrac12(\partial_yF)^2+\tfrac16F^6
    \right]_{\tau=\tau_0}.
    \label{eq:N2ClassicalEnergy}
\end{equation}
The saddle is the positive solution of
\begin{equation}
    (\partial_\tau^2+\partial_y^2)F=F^5,
    \qquad
    F(y,\tau+\beta)=F(y,\tau),
    \qquad
    F-F_\star\xrightarrow[r\to0]{}0,
    \qquad
    F\xrightarrow[|y|\to\infty]{}0.
    \label{eq:N2ThermalProblem}
\end{equation}
It is unique: the difference $w$ of two solutions obeys $\nabla^2w=q\,w$ with $q>0$ and vanishes at the line and at infinity, so it can have neither a positive maximum nor a negative minimum.
Hence $F(y,-\tau)=F(y,\tau)$.
Together with periodicity, this makes $F$ symmetric about $\tau=\beta/2$, so that $\partial_\tau F=0$ on that slice.
Choosing $\tau_0=\beta/2$, the negative term in~\eqref{eq:N2ClassicalEnergy} drops out and
\begin{equation}
    m_D
    =
    \frac{2e_0}{\sqrt3}\,k,
    \qquad
    e_0
    =
    \beta^2\int_{-\infty}^{\infty}\mathrm{d}y
    \left[
        \tfrac12(\partial_yF)^2+\tfrac16F^6
    \right]_{\tau=\beta/2} 
    >0.
    \label{eq:N2PositiveResult}
\end{equation}
Evaluating this integral numerically yields
\begin{gather}
    m_D \overset{k \to \infty}{\sim} 1.083 k
\end{gather}
The perimeter coefficient is therefore positive whatever the value of $e_0$.
The argument uses only that the line is described by a unique real saddle of scalar fields with a non-negative potential, and that it is invariant under $\tau\to-\tau$.

\section*{Acknowledgements}
 It is a pleasure to thank Daniele Artico, Zohar Komargodski,  for very useful discussions. We also thank Daniele Artico, Zohar Komargodski, Enrico Marchetto, Simone Giombi. for very useful comments on the draft. 
 The authors acknowledge the use of generative AI, in particular ChatGPT (OpenAI) and Claude (Antropic), in the preparation and editing of this manuscript. All scientific results and conclusions remain the responsibility of the human authors. In the interest of full transparency, we note that this declaration was itself written by AI, creating a self-referential loop of AI disclosure. We believe this loop is finite to all orders in perturbation theory, although a nonperturbative proof remains beyond the scope of this work.

\appendix
\section{Details on the Monte Carlo simulation}\label{app:details}

The dataset used in this work is an isotropic fixed-cutoff radius scan with
\begin{equation}
  N_\tau=8,
  \qquad
  N_x=N_y=176,
  \qquad
  \beta=8a.
\end{equation}
The full lattice therefore contains
\begin{equation}
  N_\tau N_xN_y=8\times176^2=247\,808
\end{equation}
Ising spins.  The bulk coupling is held fixed at the flat-space critical
value for every radius and every source value.

We simulate eight circular pinned magnetic loops with
\begin{equation}
  R/a=(24,30,36,42,48,54,60,64),
\end{equation}
corresponding to
\begin{equation}
  \frac{R}{\beta}=\frac{R/a}{N_\tau}
  =(3,\ 3.75,\ 4.5,\ 5.25,\ 6,\ 6.75,\ 7.5,\ 8).
  \label{eq:magnetic-loop-rho-values}
\end{equation}
At the largest radius, $R/a=64$, the loop remains $176/2-64=24$ lattice
spacings away from its nearest periodic image.

For every radius, thermodynamic integration uses the uniform source grid
\begin{equation}
  h=0,\ 0.25,\ 0.50,\ldots,8.00,
\end{equation}
namely $33$ integration nodes.  At each pair $(R,h)$ we simulate four
independent Markov chains.  Each chain has
\begin{equation}
  N_{\rm therm}=800,
  \qquad
  N_{\rm meas}=2000,
  \qquad
  N_{\rm sep}=1,
\end{equation}
where $N_{\rm sep}$ is the number of hybrid sweeps between saved
measurements.  Thus each integration node contains
\begin{equation}
  4\times2000=8000
\end{equation}
stored measurements before autocorrelation corrections.

A hybrid sweep consists of one local Metropolis sweep followed by one
ghost-Wolff cluster update for the positive magnetic source.  The actual
number of sites on the rasterized loop is recorded for every radius.  We
therefore retain both the nominal continuum variable $R/\beta$ and the
geometric diagnostic
\begin{equation}
  \frac{L_{\rm lat}}{2\pi\beta}
  =
  \frac{N_{\rm loop\ sites}}{2\pi N_\tau},
\end{equation}
which quantifies the finite-lattice-spacing difference between the smooth
circle used in the EFT and its lattice realization.

For reproducibility of the results the Monte Carlo code used for the simulations presented in this work are pubblically avaiable on \href{https://github.com/alemiscio/Spatial-Defect-MC-3D-Ising-/tree/main/}{GitHub}.
In particular the material includes both the code for reproducibility of the data and the file for the data analysis on them.

\bibliography{Bib.bib}
\bibliographystyle{JHEP}

\end{document}